%% file: reionization_v2.tex
\documentclass[useAMS,usenatbib,fleqn]{mn2e}
\usepackage{amsmath,amsfonts,amssymb}
\usepackage{booktabs}
\usepackage{graphicx}
\usepackage{pdflscape}
\usepackage{longtable}
\usepackage[normalem]{ulem}
\usepackage{footnote}
\usepackage[pdftex,breaklinks,colorlinks,citecolor=blue]{hyperref}

\pdfoutput=1
\defcitealias{Robertson:2013ji}{R13}

\title[Powering reionization]{Powering reionization: assessing the galaxy ionizing photon budget at $z < 10$}
\author[Duncan and Conselice]{Kenneth Duncan $^{1}$\thanks{E-mail:
ppxkd@nottingham.ac.uk} and Christopher J. Conselice$^{1}$ \\
$^{1}$University of Nottingham, School of Physics \& Astronomy, Nottingham NG7 2RD \\
}

\begin{document}

\date{}

\pagerange{\pageref{firstpage}--\pageref{lastpage}} \pubyear{2015}

\maketitle

\label{firstpage}
\begin{abstract}
We present a new analysis of the ionizing emissivity ($\dot{N}_{\rm{ion}}$, s$^{-1}$ Mpc$^{-3}$) for galaxies during the epoch of reionization and their potential for completing and maintaining reionization. We use extensive SED modelling -- incorporating two plausible mechanisms for the escape of Lyman continuum photon -- to explore the range and evolution of ionizing efficiencies consistent with new results on galaxy colours ($\beta$) during this epoch. We estimate $\dot{N}_{\rm{ion}}$ for the latest observations of the luminosity and star-formation rate density at $z<10$, outlining the range of emissivity histories consistent with our new model. 
Given the growing observational evidence for a UV colour-magnitude relation in high-redshift galaxies, we find that for any plausible evolution in galaxy properties, red (brighter) galaxies are less efficient at producing ionizing photons than their blue (fainter) counterparts. The assumption of a redshift and luminosity evolution in $\beta$ leads to two important conclusions. Firstly, the ionizing efficiency of galaxies naturally increases with redshift.
Secondly, for a luminosity dependent ionizing efficiency, we find that galaxies down to a rest-frame magnitude of $M_{\rm{UV}} \approx -15$ alone can potentially produce sufficient numbers of ionizing photons to maintain reionization as early as $z\sim8$ for a clumping factor of $C_{\textsc{Hii}} \leq 3$.
\end{abstract}


\section{Introduction}
At the present day, the intergalactic and interstellar medium (IGM,ISM) are known to be predominantly ionized. However, following recombination at $z\approx1100$, the baryon content of the Universe was mostly neutral. At some point in the history of the Universe, the IGM underwent a transition from this neutral phase to the ionized medium we see today, a period known as the epoch of reionization (EoR hereafter). The strongest constraints on when reionization occurred are set by observations of the Gunn-Peterson trough of distant quasars \citep{2006ARA&A..44..415F}, which indicate that by $z\approx 5.5$, the Universe was mostly ionized (with neutral fractions $\sim 10^{-4}$). 

Additionally, measurements of the total optical depth of electrons to the surface of last scattering implies that reionization should be occurring at higher redshift, towards $z \approx 10$, for models of instantaneous reionization \citep{Hinshaw:2013dd,Bennett:2013ew}. However, critical outstanding questions still remain. Firstly, when did the intergalactic hydrogen and helium complete reionization? And secondly, what were the sources of ionizing photons which powered the reionization process? Was it predominantly powered by star-forming galaxies or by active galactic nuclei/quasars?

For hydrogen reionization with its earlier completion, the rapid decline in the quasar luminosity function at high redshift \citep{Willott:2010bk,Fontanot:2012fx,Fontanot:2014jz} does suggest that star-forming galaxies are the most likely candidates for completing the bulk of reionization by $z\sim6$. The contribution from faint AGN could still however make a significant contribution to the ionizing emissivity at $z>4$ \citep{Giallongo:2015to}. Based on the optical depth constraints set by WMAP \citep{Hinshaw:2013dd} and either observed IGM emissivities at lower redshift \citep{Kuhlen:2012ka,Robertson:2013ji,Becker:2013hc}, or emissivities predicted by simulations \citep{Ciardi:2012hl}, several studies have drawn the same conclusion that faint galaxies from below the current detection limits and/or an increasing ionizing `efficiency' at higher redshift is required. Even with the lower optical depth measurement now favoured by the recent Planck analysis \citep{Collaboration:2015tp}, such assumptions are essentially still required to satisfy these criteria \citep{Robertson:2015wk}.

One of the possible mechanisms for this increasing ionizing efficiency is an evolution in the fraction of the ionizing photons able to escape their host galaxy and ionize the surrounding IGM, known as the escape fraction ($f_{\rm{esc}}$). There have been several studies designed to understand this issue, but there are still large uncertainties in what the escape fraction for galaxies is and how it evolves with redshift and other galaxy properties. In a study of $z \sim 1.3$ galaxies, \citet{Siana:2010bo} searched for Lyman-continuum photons from star forming galaxies, although no systems were detected.  After correcting for the Lyman-break and IGM attenuation the limit placed on the escape fraction is $f_{\rm esc} < 0.02$ after stacking all sources. \citet{Bridge:2010ie} find an even lower limit of $f_{\rm esc} < 0.01$ using slitless spectroscopy at $z \sim 0.7$, although one AGN in their sample is detected. However, higher escape fractions of $\sim5\%$ to $\sim20-30\%$ have been measured for galaxies at $z \sim 3$  \citep{Shapley:2006cq,Iwata:2009dy,Vanzella:2010jk,Nestor:2013kw}, consistent with the relatively high $f_{\rm esc} \sim 0.2$ expected from IGM recombination rates determined from Ly$\alpha$ forests \citep{Bolton:2007gc}. Furthermore, the average $f_{\rm{esc}}$ for galaxies at $z\sim3$ may be significantly higher than the existing measurements due to the selection biases introduced by the Lyman-break technique \citep{Cooke:2014dd}. 

It is important to bear in mind that the property which is fundamental to studies of reionization is the total number of ionizing photons which are available to ionize the intergalactic medium surrounding galaxies. Hence, while the escape fraction is a critical parameter for reionization, it must be measured or constrained in conjunction with the underlying continuum emission to which it applies. For example, an increase in $f_{\rm{esc}}$ may not have an effect on reionization if it is accompanied by a reduction in the intrinsic number of ionizing photons being produced.

As shown in \citet{Robertson:2013ji} (see also \citeauthor{Leitherer:1999jt}~\citeyear{Leitherer:1999jt}), the number of ionizing photons produced per unit UV luminosity emitted (e.g. $L_{1500\text{\AA}}$) can vary significantly as a function of the stellar population parameters such as age, metallicity and dust extinction. Thankfully, evolution or variation among the galaxy population in these parameters will not only influence the production of ionizing photons but will have an effect on other observable properties such as the UV continuum slope ($\beta$, \citeauthor{1994ApJ...429..582C}~\citeyear{1994ApJ...429..582C}).
With the advent of ultra-deep near-infrared imaging surveys such as the UDF12 \citep{Koekemoer:2013db} and CANDELS \citep{2011ApJS..197...35G,Koekemoer:2011br} surveys, observations of the UV continuum slope extending deep into the epoch of reionization are now available. Furthermore, there is now strong evidence for an evolution in $\beta$ as a function of both galaxy luminosity and redshift out to $z\sim 8$ \citep{Bouwens:2013vf,Rogers:2014bn}.

In this paper we use the latest observations of $\beta$ spanning the EoR combined with SED modelling, incorporating the physically motivated escape mechanisms to explore what constraints on the key emissivity coefficients are currently available. We also explore the consequences these constraints may have on the EoR for current observations of the star-formation rates in this epoch. In addition to the observations of in-situ star-formation through the UV luminosity functions, we also investigate whether recent measurements of the galaxy stellar mass function and stellar mass density at high redshift \citep{Duncan:2014gh,Grazian:2014vx} can provide additional useful constraints on the star-formation rates during EoR \citep{Stark:2007gi,Gonzalez:2010hm}.

In Section~\ref{sec:link}, we outline the physics and critical parameters required to link the evolution of the neutral hydrogen fraction to the production of ionizing photons. We also explore plausible physical mechanisms for the escape of Lyman continuum photons from galaxies, outlining the models explored throughout the paper. We then review the current literature constraints on the UV continuum slope, $\beta$, both as a function of redshift and galaxy luminosity. Next, in Section~\ref{sec:models}, we explore how the escape fraction, dust extinction and other stellar population parameters affect the observed $\beta$ and the coefficients relating ionizing photon production rates to observed star-formation rates and UV luminosities. In Section~\ref{sec:results}, we apply these coefficients to a range of existing observations, exploring the predicted ionizing emissivity throughout the epoch of reionization for both constant and redshift dependent conversions. We then discuss how the varying assumptions and proposed relations would impact the ionizing photon budget consistent with current observations before outlining the future prospects for improving these constraints in Section~\ref{sec:discussion}. Finally, we summarise our findings and conclusions in Section~\ref{sec:summary}.

Throughout this paper, all magnitudes are quoted in the AB system \citep{1983ApJ...266..713O}. We also assume a $\Lambda$-CDM cosmology with $H_{0} = 70$ kms$^{-1}$Mpc$^{-1}$, $\Omega_{m}=0.3$ and $\Omega_{\Lambda}=0.7$. Quoted observables (e.g. luminosity density) are expressed as actual values assuming this cosmology. We note that luminosity and luminosity based properties such as stellar masses and star-formation rates scales as $h^{-2}$, whilst densities scale as $h^{3}$.

\section{Linking reionization with observations}\label{sec:link}
\subsection{The ionizing emissivity}
The currently accepted theoretical picture of the epoch of reionization, as initially described by \citet{Madau:1999kl}, outlines the competing physical processes of ionization of neutral hydrogen by Lyman continuum photons and recombination of free electrons and protons. The transition from a neutral Universe to a fully ionized one can be described by the differential equation:
\begin{equation}
\dot{Q}_\textsc{Hii} = \frac{\dot{N}_{\rm{ion}}}{\left \langle n_{\textsc{H}} \right \rangle} - \frac{{Q}_\textsc{Hii}}{\left \langle t_{\rm{rec}} \right \rangle}
\end{equation}
\noindent where ${Q}_\textsc{Hii}$ is the dimensionless filling factor of ionized hydrogen (such that ${Q}_\textsc{Hii} = 1$ for a completely ionized Universe) and $\dot{N}_{\rm{ion}}$ is the comoving ionizing photon production rate (s$^{-1}$ Mpc$^{-3}$) or ionizing emissivity. The comoving density of hydrogen atoms, $\left \langle n_{\textsc{H}} \right \rangle$, and average recombination time $\left \langle t_{\rm{rec}} \right \rangle$ are redshift dependent and is dependent on the primordial Helium abundance, IGM temperature and crucially the inhomogeneity of the IGM, parametrised as the so-called clumping factor $C_{\textsc{Hii}} \equiv \left \langle n^{2}_{\textsc{H}} \right \rangle / \left \langle n_{\textsc{H}} \right \rangle^{2}$\citep{Pawlik:2009id}. 
We refer the reader to \citet{Madau:1999kl}, \citet{Kuhlen:2012ka} and \citet{2010Natur.468...49R,Robertson:2013ji} for full details on these parameters and the assumptions associated. 

In this paper we will concentrate on $\dot{N}_{\rm{ion}}$ and the production of Lyman continuum photons by star-forming galaxies. The link between the observable properties of galaxies and the ionizing photon rate, $\dot{N}_{\rm{ion}}$ can be parametrised as
\begin{equation}\label{eq:Nion_UV}
\dot{N}_{\rm{ion}} = f_{\rm{esc}}\xi_{\rm{ion}}\rho_{\rm{UV}}
\end{equation}
\noindent following the notation of \citet{Robertson:2013ji} (R13 hereafter), where $\rho_{\rm{UV}}$ is the observed UV (1500\AA) luminosity density (in erg s$^{-1}$ Hz$^{-1}$ Mpc$^{-3}$), $\xi_{\rm{ion}}$ the number of ionizing photons produced per unit UV luminosity (erg$^{-1}$ Hz) and $f_{\rm{esc}}$ the fraction of those photons which escape a host galaxy into the surrounding IGM. Alternatively, $\dot{N}_{\rm{ion}}$ can be considered in terms of the star-formation rate density, $\rho_{\rm{SFR}}$ (M$_{\odot}$ yr$^{-1}$ Mpc$^{-3}$),
\begin{equation}\label{eq:Nion_SFR}
    \dot{N}_{\rm{ion}} = f_{\rm{esc}} \kappa_{\rm{ion}} \rho_{\rm{SFR}}
\end{equation}

\noindent where $\kappa_{\rm{ion}}$ (s$^{-1}$ M$_{\odot}^{-1}$ yr) is the ionizing photon production-rate per unit star-formation ($\zeta_{Q}$ in the notation used in \citet{2010Natur.468...49R}). In addition to observing $\rho_{\rm{UV}}$ or $\rho_{\rm{SFR}}$ during the EoR, accurately knowing $f_{\rm{esc}}$, $\xi_{\rm{ion}}$ and $\kappa_{\rm{ion}}$ is therefore critical to estimating the total ionizing emissivity independent of how well we are able to measure the UV luminosity or star-formation rate densities.

As the production of $LyC$ photons is dominated by young UV-bright stars, the rate of ionizing photons is therefore highly dependent on the age of the underlying stellar population and the recent star-formation within the galaxy population. Physically motivated values of $\xi_{\rm{ion}}$ (or its equivalent coefficient in other notation) can be estimated from stellar population models based on plausible assumptions on the properties of high-redshift galaxies \citep{Bolton:2007gc,Ouchi:2009jd,Kuhlen:2012ka}. 

However, with observations of the UV continuum slope, $\beta$ \citep{1994ApJ...429..582C}, now extending deep into the epoch of reionization, limited spectral information is now available for a large sample of galaxies. Despite the many degeneracies in $\beta$ (see later discussion in Section~\ref{sec:models}), it is now possible to place some constraints on whether the assumptions made are plausible. In \citetalias{Robertson:2013ji}, values of $\xi_{\rm{ion}}$ are explored for a range of stellar population parameters relative to the $\beta$ observations of \citet{Dunlop:2013kp}. Based on the range of values consistent with the observed values of $\beta \approx -2$, \citeauthor{Robertson:2013ji} choose a physically motivated value of $\log_{10}\xi_{\rm{ion}} = 25.2$. 

Typically, a constant $f_{\rm{esc}}$ is applied to all galaxies in addition to the estimated or assumed values of $\xi_{\rm{ion}}$/$\kappa_{\rm{ion}}$ \citep{Ouchi:2009jd,Finkelstein:2012hr,Robertson:2013ji}, motivated in part by our lack of understanding of the redshift or halo mass dependence of $f_{\rm{esc}}$. However, applying a constant $f_{\rm{esc}}$ does not take into account exactly how the Lyman continuum photons escape the galaxy, what effect the different escape mechanism might have on the observed galaxy colours, and how that might alter the assumed $\xi_{\rm{ion}}$/$\kappa_{\rm{ion}}$ based on $\beta$. 

\subsection{Mechanisms for Lyman continuum escape}\label{sec:escape_mechanisms}

\begin{figure*}
\centering
  \includegraphics[width=0.42\textwidth]{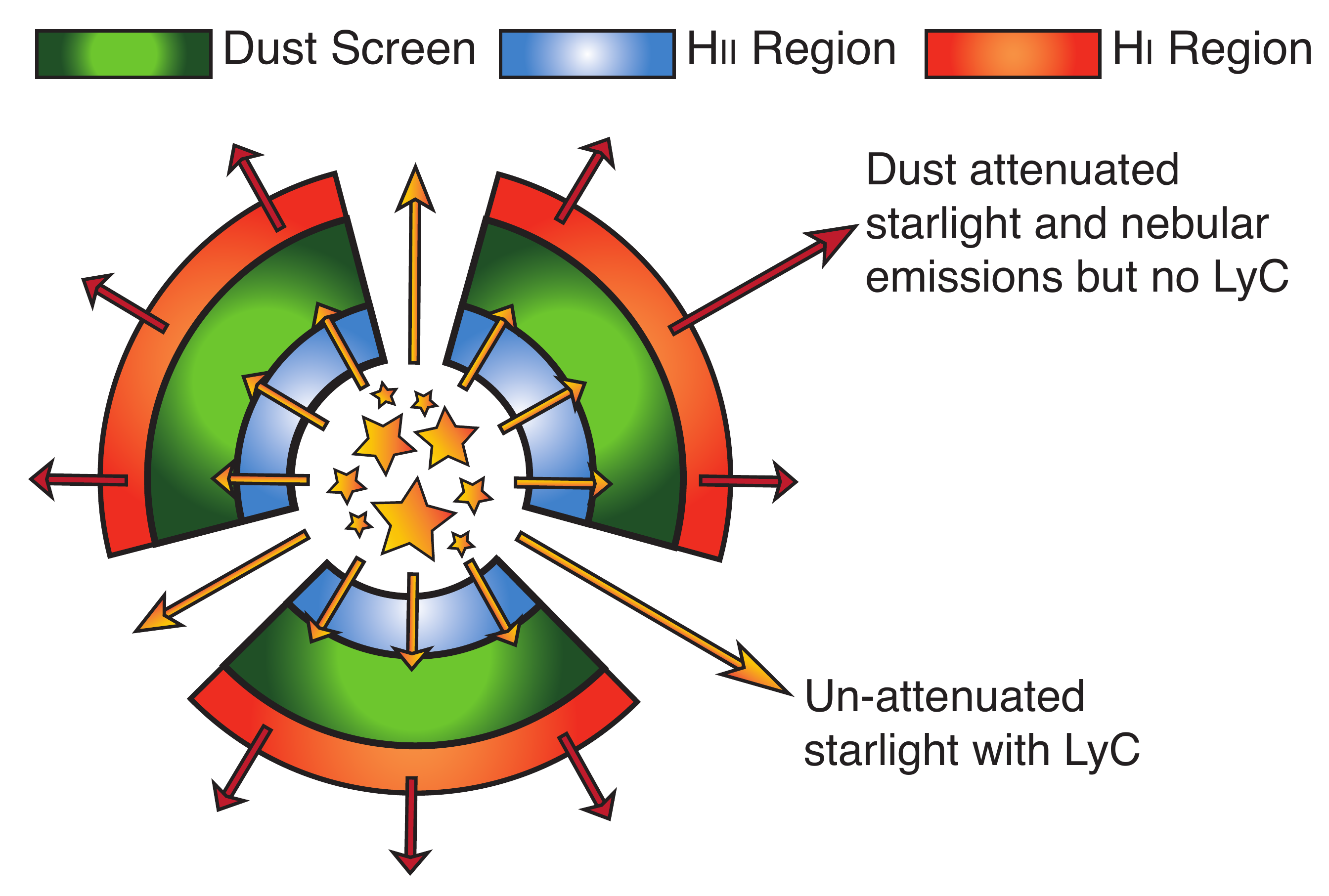}
  \quad \quad \quad \quad \quad
  \includegraphics[width=0.42\textwidth]{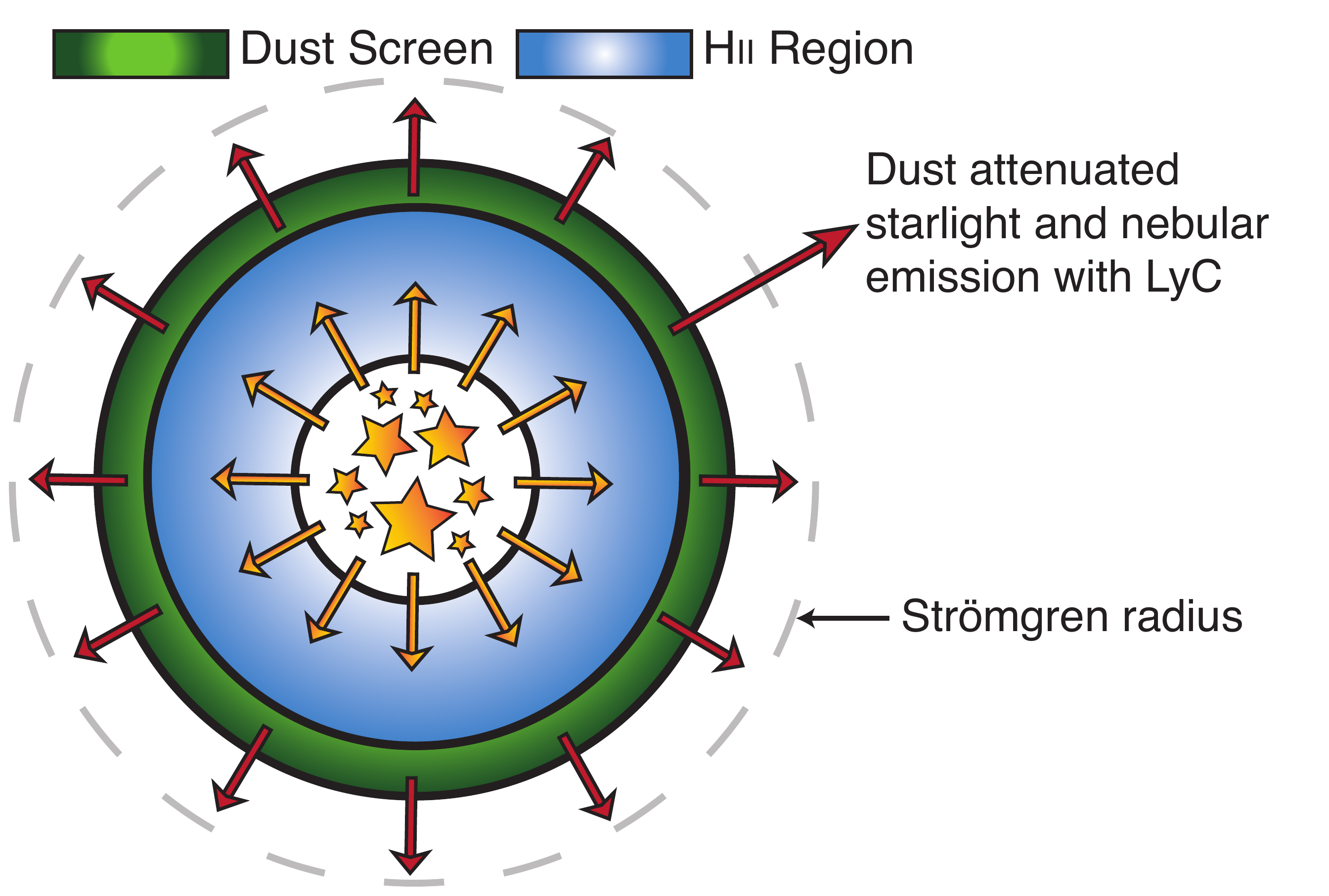}

  \caption{Schematic cartoon illustrations of the Lyman continuum escape mechanisms outlined in Section~\ref{sec:link}. For both models, the stars represent a central galaxy surrounded by a {\sc Hii} ionization region, dust is distributed in an outer dust-screen. Left: An ionization bounded nebula with holes (sometimes referred to as the 'picket fence model') in which LyC escapes through holes in the ISM. Right: a density bounded nebula where LyC is able to escape due to the incomplete Str{\"o}mgren sphere formed when the galaxy depletes its supply of neutral hydrogen.}
  \label{fig:mechanisms}
\end{figure*}

In \citet{Zackrisson:2013iz}, detailed SSP and photo-dissociation models were used to explore how future observations of $\beta$ and H$\alpha$ equivalent-width with the James Webb Space Telescope (JWST) could be used to constrain the escape fraction for two different Lyman continuum escape mechanisms (see Fig.~8 of \citet{Zackrisson:2013iz}). However, it may already be possible to rule out significant parts of the $f_{\rm{esc}}$ and dust extinction parameter space using current constraints on $\beta$ and other galaxy properties. To estimate the existing constraints on $f_{\rm{esc}}$, $\xi_{\rm{ion}}$, $\kappa_{\rm{ion}}$ and their respective products, we combine the approaches of \citet{Zackrisson:2013iz} and \citet{2010Natur.468...49R, Robertson:2013ji}. To do this, we model $\beta$, $\xi_{\rm{ion}}$ and $\kappa_{\rm{ion}}$ as a function of $f_{\rm{esc}}$ for the two models of \citet{Zackrisson:2013iz}. The components and geometry of these two models are illustrated in Fig.~\ref{fig:mechanisms}.

In the first model, model A hereafter (Fig.~\ref{fig:mechanisms} left) and dubbed `ionization bounded nebula with holes' by \citeauthor{Zackrisson:2013iz}, Lyman continuum photons along with unattenuated starlight are able to escape through low-density holes in the neutral ISM. In this model, the escape fraction is determined by the total covering fraction of the neutral ISM. Under the assumption that these holes are small and evenly distributed, the observed galaxy SED (averaged across the galaxy as in the case of photometry) would then be a combination of the unattenuated starlight from holes and the dust reddened starlight and nebular emission from the H{\sc i} enshrouded regions.

A second model, model B hereafter (Fig.~\ref{fig:mechanisms} right) corresponds to the `density bounded nebula' of \citet{Zackrisson:2013iz}. This model could occur when the local supply of H{\sc i} is exhausted before a complete Str{\"o}mgren sphere can form, allowing Lyman continuum photons to escape into the surrounding ISM. The fraction of LyC photons which can escape the nebular region is determined by the fraction of the full Str{\"o}mgren radius at which the nebular region is truncated. The total escape fraction is then also dependent on the optical depth of the surrounding dust screen. Of these two mechanisms, the former (Model A: ionization bounded nebula with holes) is the model which most closely represents the physics predicted by full radiation hydrodynamical models of dwarf galaxies. 

In \citet{Wise:2009fn}, it was found that Lyman continuum radiation preferentially escaped through channels with low column densities, produced by radiative feedback from massive stars. The resulting distribution of LyC escape fraction within a galaxy is highly anisotropic and varies significantly between different orientations. Evidence for such an anisotropic escape mechanism has also been found recently by \citet{Zastrow:2013bg}, who find optically thin ionization cones through which LyC can escape in nearby dwarf starbursts. Similarly, \citet{Borthakur:2014bz} find a potential high-redshift galaxy analog at $z\sim0.2$ with evidence for LyC leakage through holes in the surrounding neutral gas with an escape fraction as high as $f_{\rm{esc}} \approx 0.2$ (21\%). 

However, this value represents the optimum case in which there is no dust in or around the low-density channels (corresponding to Model A). For the same system, when \citet{Borthakur:2014bz} include dust in the low-density channels, the corresponding total LyC escape fraction is reduced to $\approx 1\%$. The two models explored in this work represent the two extremes of how dust extinction will effect the escaping Lyman continuum for toy models such as these, the dust-included estimates of \citet{Borthakur:2014bz} therefore represent a system which lies somewhere between Models A and B. 

A potential third mechanism for Lyman continuum escape was posited by \citet{Conroy:2012fc}, whereby `runaway' OB stars which have traveled outside the galaxy centre can contribute a significant amount to the LyC emitted into the surrounding IGM. For high-redshift galaxies with significantly smaller radii than local galaxies, massive stars with large velocities could venture up to 1 kpc away from their initial origin into regions with low column density. \citet{Conroy:2012fc} estimate that these stars could in fact contribute $50 - 90\%$ of the escaping ionizing radiation. In contrast, recent work by \citet{Kimm:2014gv} finds that when runaway stars are included into their models of Lyman continuum escape, the time average escape fraction only increases by $\sim20\%$. Given the additional complications in modelling the relevant observational properties and their relatively small effect, we neglect the contribution of runaway stars in the subsequent analysis. 

In Section~\ref{sec:models}, we describe how we model both the observable ($\beta$) and unobservable ($\xi_{\rm{ion}}$,$\kappa_{\rm{ion}}$) properties for both model A and model B. But first, we examine the existing observations on the evolution of $\beta$ into the epoch of reionization.

\subsection{Observed UV Continuum Slopes}\label{sec:betas}
\begin{figure}
  \includegraphics[width=0.48\textwidth]{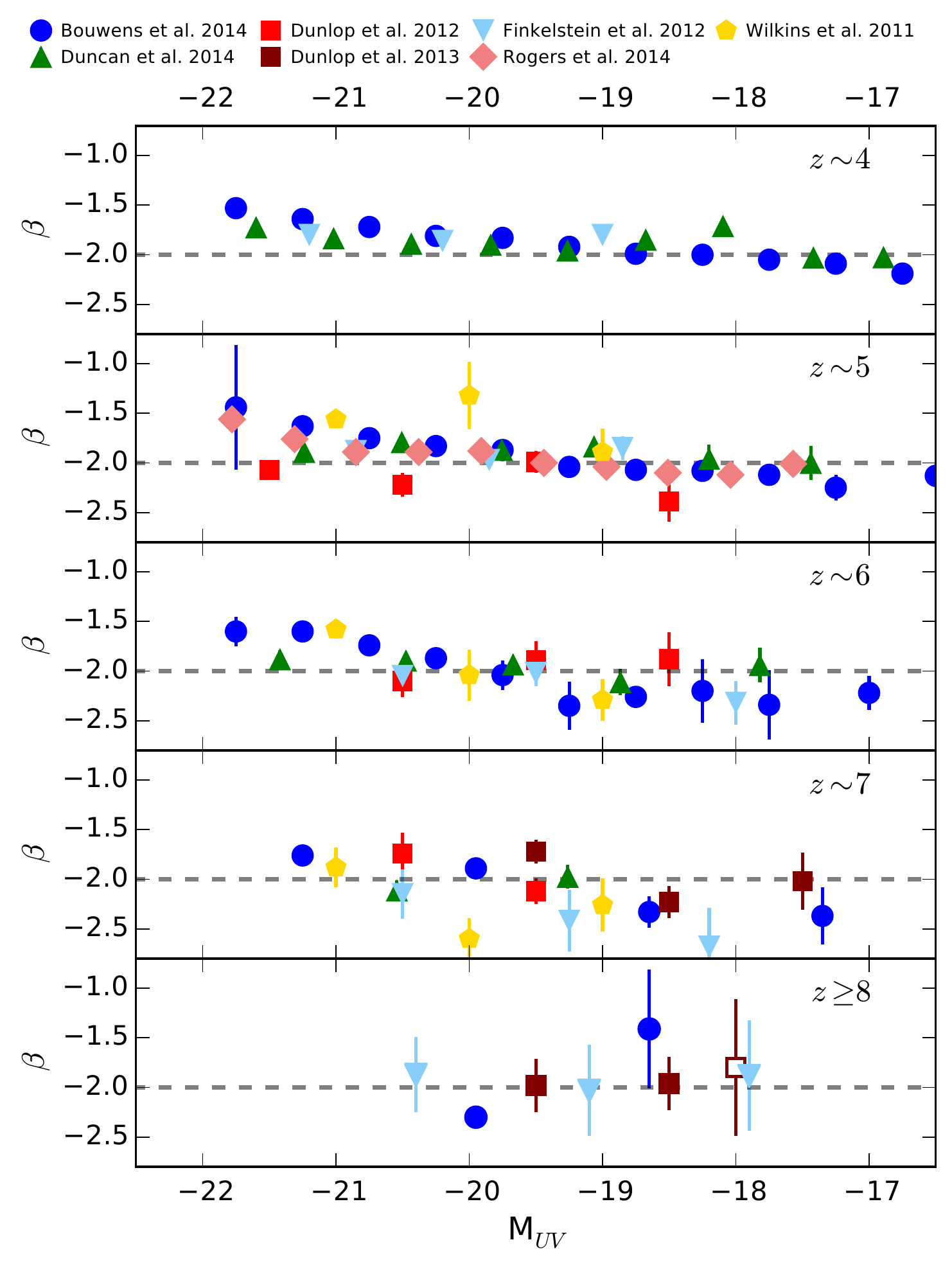}
  \caption{Observed average values of the UV continuum slopes $\beta$ as a function of rest-frame UV magnitude, $M_{\rm{UV}}$, from \citet{Wilkins:2011fs}, \citet{Dunlop:2011jl,Dunlop:2013kp}, \citet{2012ApJ...756..164F}, \citet{Bouwens:2013vf}, \citet{Duncan:2014gh} and \citet{Rogers:2014bn} at redshifts $z\sim4$, 5, 6, 7 and $8-9$. In the bottom panel, filled symbols show the average for $8\sim7$ samples while the open symbol shows the averages for $z\geq 9$ (see respective papers for sample details and redshift ranges).}
  \label{fig:beta_z}
\end{figure}

In Fig.~\ref{fig:beta_z}, we show a compilation of recent results in the literature on the observed UV slope, $\beta$, as a function of both redshift and rest-frame UV magnitude, $M_{\rm{UV}}$ \citep{Dunlop:2011jl,Dunlop:2013kp,Wilkins:2011fs,2012ApJ...756..164F,Bouwens:2013vf,Duncan:2014gh,Rogers:2014bn}. Disagreement between past observations on the existence or steepness of a color-magnitude relation (cf. \citet{Dunlop:2011jl} and \citet{Bouwens:2011tj}) have recently been reconciled by \citet{Bouwens:2013vf} after addressing systematics in the selection and photometry between different studies. \citet{Bouwens:2013vf} find a clear colour-magnitude relation (CMR) with bluer UV-slopes at lower luminosities, the relation is also found to evolve with bluer $\beta$'s at high redshift (blue circles in Fig.~\ref{fig:beta_z}). The existence of a strong colour-magnitude relation has also been confirmed by \citet{Rogers:2014bn} at $z\sim5$ for a sample of even greater dynamic range (pink diamonds in Fig.~\ref{fig:beta_z}), measuring a CMR slope and intercept within error of the measured $z\sim5$ values of \citet{Bouwens:2013vf}.

\begin{table}
\begin{minipage}{0.475\textwidth}
\centering
  \caption{Bayesian Information Criterion (BIC) for the assumption of either a colour-magnitude relation or a constant $\beta$, $\Delta$BIC is defined as BIC$_{\rm{const}} -$BIC$_{\rm{slope}}$. For each dataset, we also show the best-fit model parameters and corresponding 1-$\sigma$ errors for the model with lowest BIC.}	
	\begin{tabular}{cccc}\label{tab:BIC_muv}
    	Redshift	& BIC$_{\rm{slope}}$ & BIC$_{\rm{const}}$ & $\Delta$BIC \\
    \hline
	$z\sim7$  & 35.2 \footnote[1]{\( \beta(M_{\rm{UV}}) = -2.05 \pm 0.04 - 0.13 \pm 0.04\times (M_{\rm{UV}} + 19.5)\)}  & 45.0 & 9.8 \\
	$z\sim8$ &  5.8  & 4.2 \footnote[2]{\( \beta = -2.00 \pm 0.11 \)} & -1.6 \\
  \end{tabular}
  \end{minipage}

\end{table}

While the evidence for a colour-magnitude relation at $z\leq6$ is quite strong, during the crucial period of reionization at $z\sim7$ the limited dynamic range and samples sizes of existing observations means that similar conclusions are less obvious. Given the importance of this period in reionization and also the importance the assumption of a colour-magnitude relation has on the conclusions of this study, it is pertinent to critically assess which model is statistically favoured by the existing observations. To assess the statistical evidence for a colour-magnitude relation at $z\sim7$, we calculate the Bayesian information criterion (BIC, \citeauthor{Schwarz:1978uv}~\citeyear{Schwarz:1978uv}), defined as
\begin{equation}
	BIC = 	-2 \ln\mathcal{L}_{\rm{max}} + k \ln N,
\end{equation}
where $\mathcal{L}_{\rm{max}}$ is the maximised value of the likelihood function for the model in question ($-2 \ln \mathcal{L}_{\rm{max}} \equiv \chi^{2}_{\rm{min}}$ under the assumption of gaussian errors), $k$ the number of parameters in said model and $N$ the number of data-points being fitted. Values of $\Delta$BIC greater than 2 are positive evidence against the model with higher BIC, whilst values greater than 6 (10) are strong (very strong) evidence against. The model fits were performed using the MCMC implementation of \citet{ForemanMackey:2013io} assuming a flat-prior. The resulting BIC for the constant $\beta$ and linear $M_{\rm{UV}}$-dependent models are shown in Table~\ref{tab:BIC_muv}. For the $z\sim7$ observations, a color-magnitude relation is strongly favoured over a constant $\beta$ with a best-fit model of \( -2.05\pm0.04 -0.13\pm0.04(M_{\rm{UV}} + 19.5)\). At $z\sim8$, while the assumption of a constant $\beta$ provides a better fit ($\beta = -2.00 \pm 0.11$), no model is strongly preferred over the other.

Past studies into the ionizing emissivity during the EoR have often used a fixed average $\beta = -2$ to motivate or constrain $\xi_{\rm{ion}}$, e.g. \citet{Bolton:2007gc}, \citet{Ouchi:2009jd}, \citet{2010Natur.468...49R,Robertson:2013ji} and \citet{Kuhlen:2012ka}. Although we now find good evidence for a colour-magnitude relation during this epoch, it does not necessarily make the assumption of a constant $\beta$ invalid, as we must take into account the colours of the galaxies which dominate the SFR or luminosity density. In order to estimate an average $\beta$ which takes into account the corresponding number densities and galaxy luminosities, we calculate $\left \langle \beta  \right \rangle_{\rho_{\rm{UV}}}$, the average $\beta$ weighted by the contribution to the total UV luminosity density:
\begin{equation}
\left \langle \beta  \right \rangle_{\rho_{\rm{UV}}} = \frac{\int_{L_{\rm{min}}}^{\infty}  L_{\rm{UV}}(m) \times \phi(m) \times \left \langle \beta  \right \rangle (m)}{\int_{L_{\rm{min}}}^{\infty} L_{\rm{UV}}(m) \times \phi(m)}
\end{equation}
\noindent where $L_{\rm{UV}}(m)$, $\phi(m)$ and $\left \langle \beta  \right \rangle (m)$ are the luminosity, number density and average $\beta$ at the rest-frame UV magnitude, $m$, respectively. We choose a lower limit of $L_{\rm{UV}} \equiv M_{\rm{UV}} = -17$, corresponding to the approximate limiting magnitude of the deepest surveys at $z\geq6$. For the discrete bins in which $\left \langle \beta  \right \rangle (m)$ is calculated, $\left \langle \beta  \right \rangle_{m}$, this becomes a sum over the bins of absolute magnitude, $k$, brighter than our lower limit $M_{\rm{UV}} = -17$:
\begin{equation}
\left \langle \beta  \right \rangle_{\rho_{\rm{UV}}} = \frac{\sum_{k}^{ }  L_{\rm{UV}, \mathit{k}} \times \phi_{k} \times \left \langle \beta  \right \rangle_{k}}{\sum_{k}^{ } L_{\rm{UV},  \mathit{k}} \times \phi_{k}.}
\end{equation}
\noindent The number density for a given rest-frame magnitude bin, $\phi_{k}$, is calculated from the best-fitting UV luminosity functions of \citet{Bouwens:2014tx} at the corresponding redshift. We use the same luminosity function at each redshift for all of the observations for consistency. We note that given the relatively good agreement between estimates given their errors, the use of differing luminosity function estimates would have minimal effect on the calculated values. 

We estimate errors on $\left \langle \beta  \right \rangle_{\rho_{\rm{UV}}}$ through a simple Monte Carlo simulation, whereby $\left \langle \beta  \right \rangle_{m}$ and the best-fitting \citet{Schechter:1976gl} parameters used to calculate $\phi_{m}$ are perturbed by the quoted errors (making use of the full covariance measured by \citeauthor{Bouwens:2014tx}~\citeyear{Bouwens:2014tx}), this is repeated $10^{4}$ times. $\left \langle \beta  \right \rangle_{\rho_{\rm{UV}}}$ and error are then taken as the median and $1\sigma$ range of the resulting distribution.

\begin{table}
\begin{minipage}{0.475\textwidth}
\centering
  \caption{Bayesian Information Criterion (BIC) for the assumption of a redshift dependent or constant $\left \langle \beta  \right \rangle_{\rho_{\rm{UV}}}$, $\Delta$BIC is defined as BIC$_{\rm{const}} -$BIC$_{z}$. For each dataset, we also show the best-fit model parameters and corresponding 1-$\sigma$ errors for the model with lowest BIC.}	
	\begin{tabular}{cccc}\label{tab:BIC_betaz}
    	Redshift range	& BIC$_{z}$ & BIC$_{\rm{const}}$ & $\Delta$BIC \\
    \hline
	$4 \lesssim z \lesssim 8$  & 44.0   \footnote[1]{\(\left \langle \beta \right \rangle_{\rho_{\rm{UV}}}(z) = -1.59 \pm 0.05 - 0.07 \pm 0.01\times z.\)} & 76.9  & 32.8 \\
	$5 \lesssim z \lesssim 8$ & 41.0   \footnote[2]{\(\left \langle \beta \right \rangle_{\rho_{\rm{UV}}}(z) = -1.61 \pm 0.12 - 0.07 \pm 0.02\times z.\)} & 46.2 & 5.2 \\
  \end{tabular}
  \end{minipage}
\end{table}

\begin{figure}
  \includegraphics[width=0.48\textwidth]{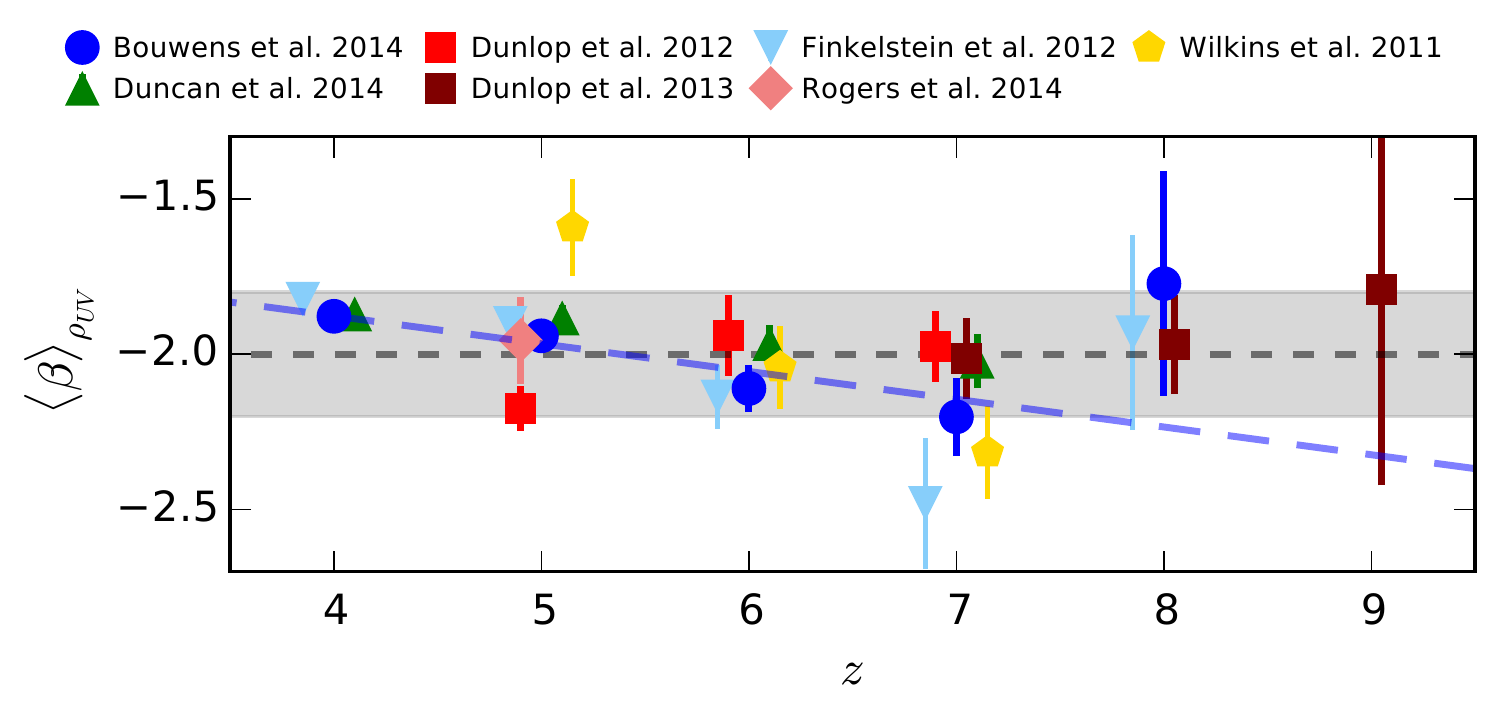}
  \caption{Luminosity-weighted average $\beta$, $\left \langle \beta  \right \rangle_{\rho_{\rm{UV}}}$, as a function of redshift for the $M_{\rm{UV}}-\beta$ observations shown in Fig.~\ref{fig:beta_z}. The grey shaded region covers the range $-2.2 < \beta < -1.8$ and the blue long-dashed line shows our parametrisation of $\left \langle \beta \right \rangle_{\rho_{\rm{UV}}}$ vs $z$ based on the observations of \citet{Bouwens:2013vf} (see Equation~\ref{eq:beta_muv_fit}).}
  \label{fig:beta_weightedavg}
\end{figure}

Fig.~\ref{fig:beta_weightedavg} shows the calculated $\left \langle \beta  \right \rangle_{\rho_{\rm{UV}}}$ and corresponding errors for each of the samples shown in Fig.~\ref{fig:beta_z}. Overall, we can see that $\beta \approx -2$ is still a valid choice for a fiducial value of $\beta$ during the epoch of reionization ($z > 6$) based on all of the existing observations. However, for several of the sets of observations there is evidence for a potential evolution at $z < 7$ \citep{Wilkins:2011fs,Finkelstein:2012hr,Bouwens:2013vf,Duncan:2014gh}. 

To determine whether there is statistical evidence for a redshift evolution in $\left \langle \beta  \right \rangle_{\rho_{\rm{UV}}}$, we again calculate the corresponding Bayesian information criteria to assess the relative merits of each model, assuming a simple linear evolution for the redshift dependent model. The resulting BIC and $\Delta$BIC are listed in Table~\ref{tab:BIC_betaz}. Based on the observations in the redshift range $4 \lesssim z \lesssim 8$, there is very strong evidence for a redshift-dependent $\left \langle \beta  \right \rangle_{\rho_{\rm{UV}}}$ over one which is constant. Because the model fits may be dominated by the $z\sim 4$ observations and their significantly smaller errors, we also calculate the fits using only the $5 \lesssim z \lesssim 8$ data. At $z \gtrsim 5$, the collective observations do still favour redshift evolution, but the statistical significance is notably reduced.

Throughout this work we choose to use the $\beta$ observations of \citet{Bouwens:2013vf} for both $\beta$ vs $M_{\rm{UV}}$ and $\left \langle \beta  \right \rangle_{\rm{UV}}$ as the basis of our analysis. This choice is motivated by the fact that these observations represent the largest samples studied (both in number and dynamic range) and due to the careful minimisation of the potential systematic errors are likely the least-biased observations. For this set of observations, there is strong evidence for evolution in $\left \langle \beta  \right \rangle_{\rm{UV}}$ at $z < 7$. Specifically, from $z\sim4$ to $z\sim7$, $\left \langle \beta  \right \rangle_{\rm{UV}}$ steepens considerably from  $-1.9\pm0.02$  to $-2.21\pm0.14$. Parametrising the \citet{Bouwens:2013vf} observations with a simple linear relation , we find:
\begin{equation}\label{eq:beta_muv_fit}
    \left \langle \beta \right \rangle_{\rho_{\rm{UV}}}(z) = -1.52 \pm 0.11 - 0.09 \pm 0.02\times z.
\end{equation}
Based on this fit we predict a $\left \langle \beta \right \rangle_{\rho_{\rm{UV}}} = -2.24$ for $z\sim8$. Whilst this is significantly bluer than that based on the existing observations at $z\sim8$, it is comparable to the average $\beta$ measured for fainter galaxies in the lower redshift samples and represents a reasonable extrapolation. We note that at $z\approx 8$, the $\left \langle \beta \right \rangle_{\rho_{\rm{UV}}}$ changes significantly depending on the choice of average due to the significantly smaller samples observable and large scatter in the faintest bin. For example, using the bi-weight means recommended by \citet{Bouwens:2013vf}, $\left \langle \beta \right \rangle_{\rho_{\rm{UV}}} = -1.74$ based on their observations. Re-calculating using the inverse-weighted means of this same sample gives $\left \langle \beta  \right \rangle_{\rho_{\rm{UV}}} = -2.1$, in better agreement with the observed trend at $z < 8$. 

While there is now good agreement on the existence and slope of the colour-magnitude relation between independent studies (cf.  \citeauthor{Bouwens:2013vf}~\citeyear{Bouwens:2013vf} and \citeauthor{Rogers:2014bn}~\citeyear{Rogers:2014bn}), what is less well understood is the intrinsic scatter in the CMR and whether it is luminosity dependent. Currently, the most extensive study of the intrinsic scatter is that of \citet{Rogers:2014bn}, who found that the intrinsic scatter in $\beta$ is significantly larger for bright galaxies. They also find an apparent lower limit (25th percentile) of $\beta = -2.1$ which varies little with galaxy luminosity whilst the corresponding 75th percentiles increase significantly from fainter to brighter galaxies. Such a scenario implies that galaxies with $\beta \leq -2.5$ should be extremely rare at high-redshift, even though such galaxies are observed locally and at intermediate redshifts \citep{Stark:2014fa}. Without a better understanding of the causes of the intrinsic scatter in $\beta$ and the underlying stellar populations it is difficult to predict the expected numbers of such galaxies during this epoch.

\citeauthor{Rogers:2014bn} interpret the intrinsic scatter as consistent with two simple scenarios: 1) the scatter is due to galaxy orientation, or 2) that brighter galaxies have more stochastic star-formation histories and the $\beta$ variation is a result of observing galaxies at different points in the duty cycle of star-formation. However, this second scenario is contrary to the theoretical predictions of \citet{Dayal:2013jm}, whereby fainter low-mass galaxies have more stochastic star-formation histories due to the greater effect of feedback shutting down star-formation in low-mass haloes.
 
After the discovery that high-redshift galaxies exhibit significant UV emission lines by \citet{Stark:2014vc} (specifically {\sc Ciii]} at 1909$\rm{\AA}$), it is worth asking if the presence of such far-UV emission lines can systematically affect measurements of the UV slope to the same degree which optical emission lines can affect age estimates and stellar masses. In \citet{Stark:2014fa}, the authors find that the fitting of $\beta$ is not affected by the presence of UV emission lines in a sample of young low-mass galaxies at $z\sim2$. The same is true for galaxies out to $z \lesssim 6$, where $\beta$ is typically measured by fitting a power-law to three or more filters \citep{Bouwens:2013vf}. However, at $z \geq 7$ where $\beta$ must be measured using a single colour, the effect of UV emission line contamination in one of the filters is more significant. For a {\sc Ciii]} equivalent width of $13.5\rm{\AA}$ (the highest observed in the \citet{Stark:2014fa} sample at $z\sim2$) could result in a measured $\beta$ which is too red by $\Delta\beta \approx 0.18$ relative to the intrinsic slope based on the method outlined in Section~\ref{sec:models}. 

Given the limited samples of $z\gtrsim 6$ galaxies with UV emission line detections, fully quantifying the effects of the emission line contamination on $\beta$ observations is not possible at this time. As such, in this work we do not include UV emission lines in our simulated SEDs or attempt to correct for their effects on the observed $\beta$s in Fig.~\ref{fig:beta_weightedavg}. We do caution that despite the significant improvement on  $\beta$ measurements at high redshift, there may still be unquantified systematics when interpreting the UV slope during the EoR.

\section{Modelling $\beta$, $\xi_{\lowercase{ion}}$ and $\kappa_{\lowercase{ion}}$} \label{sec:models}
To model the apparent $\beta$'s and corresponding emissivity coefficients for our two Lyman continuum escape models, we make use of composite stellar population models from \citet{Bruzual:2003ckb} (BC03). Using a stellar population synthesis code, we are able to calculate the full spectral energy distribution for a stellar population of the desired age, star-formation history, metallicity and dust extinction. Our code allows for any single-parameter star-formation history (e.g. exponential decline, power-law, truncated or `delayed' star-formation models), and a range of dust extinction models. The models also allow for the inclusion of nebular emission (both line and continuum emission) proportional to the $LyC$ photon rate, full details of which can be found in \citet{Duncan:2014gh}.

For each resulting SED with known star-formation rate (SFR), we calculate the UV luminosity by convolving the SED with a top-hat filter of width $100\rm{\AA}$ centred around $1500\rm{\AA}$, as is standard practice for such studies (e.g. \citeauthor{Finkelstein:2012hr}~\citeyear{Finkelstein:2012hr}, \citeauthor{McLure:2013hh}~\citeyear{McLure:2013hh}). To measure \(\beta \), each SED is redshifted to \(z \sim 7 \) and convolved with the WFC3 F125W and F160W filter responses (hereafter \(J_{125} \) and \(H_{160}\) respectively). We then calculate \( \beta \) as: 
\begin{equation}\label{eq:beta_calc}
    \beta = 4.43(J_{125} - H_{160}) - 2. 
\end{equation}
as in \citet{Dunlop:2013kp}. This method is directly comparable to how the majority of the high redshift observations were made and should allow for direct comparison when interpreting the observations with these models. Using comparable colours at $z\sim5$ and $z\sim6$ or different combinations of filters has minimal systematic effect on the calculated values of $\beta$ (see \citeauthor{Dunlop:2011jl}~\citeyear{Dunlop:2011jl} and Appendix of \citet{}\citeauthor{2012ApJ...754...83B}~\citeyear{2012ApJ...754...83B}).

The $LyC$ flux from these models is calculated before and after the applied absorption by gas (for nebular emission) and dust. We are therefore able to calculate the total escape fraction of $LyC$ photons for a given stellar population. Using these values, it is therefore relatively straight-forward to link the observed $\beta$ distribution of high-redshift galaxies with the distribution of predicted $\kappa_{\rm{ion}}$ or $L_{\rm{UV}}$ per unit SFR and the corresponding $f_{esc,tot}$.

For the ionization-bounded nebula with holes model (Model A), the `observed' SED is a weighted (proportional to $f_{\rm{esc}}$) sum of the un-attenuated starlight escaping through holes and the attenuated starlight and nebular emission from the H {\sc ii} and dust enclosed region. The two SED components are weighted proportional to the covering fraction ($\equiv 1 - f_{\rm{esc}}$) of the H {\sc ii} and dust region. The resulting observable quantities are effectively the average over all possible viewing angles, as would be expected for a large sample of randomly aligned galaxies.

In the case of Model B, the density bounded nebula, Lyman continuum emission from the underlying stellar spectrum is partially absorbed by the surrounding truncated Str\"{o}mgren sphere with an escape fraction $f_{esc,neb}$. The remaining Lyman continuum photons along with the UV-optical starlight and nebular emission are then attenuated by the surrounding dust shell according to the chosen dust attenuation law. The total escape fraction for this model is therefore
\begin{equation}
f_{\rm{esc}} = 10^{0.4\times A(LyC)}f_{\rm{esc,neb}}
\end{equation}
where $A(LyC)$ is the magnitude of dust extinction for Lyman continuum photons and is highly dependent on the choice of attenuation curve (see Section~\ref{sec:models_dust}).

\subsection{Modelling assumptions: current constraints on stellar populations at $z>3$}\label{sec:assumptions}
Although there are now good constraints on both the UV luminosity function and UV continuum slope at high redshift, both of these values suffer strong degeneracies with respect to many stellar population parameters. As such, constraints on $f_{\rm{esc}}$, $\xi_{\rm{ion}}$ and $\kappa_{\rm{ion}}$ still requires some assumptions or plausible limits set on the range of some parameters. In this section we outline the existing constraints on the relevant stellar population properties at high-redshift, discuss what assumptions we make in our subsequent analysis, and explore the systematic effects of variations in these assumptions.  

\subsubsection{Star-formation history}\label{sec:sfh}
Typically, star formation histories are parameterised as exponential models,
\begin{equation}
    \rho_{\rm{SFR}}(t) \propto \exp(-\frac{t - t_{\rm{f}}}{\tau}),
\end{equation}
where $\tau$ is the characteristic timescale and can be negative or positive (for increasing or decreasing SFH respectively). Or, alternatively as a power-law, following
\begin{equation}
    \rho_{\rm{SFR}}(t) \propto ({t - t_{\rm{f}}})^{\alpha}.
\end{equation}
The star-formation history (SFH) of high-redshift galaxies, $SFR(t)$, is still very poorly constrained due to the limited rest-frame wavelengths available for SED fitting or spectroscopy. For large samples of both intermediate and high redshift galaxies, it has been found that rising SFHs produce better SED fits to the observed photometry \citep{Maraston:2010dl,Lee:2014in}. However, reliably constraining the characteristic timescales, $\tau$ or $\alpha$, for individual galaxies at $z > 2$ is not possible for all but the brightest sources.

Using a comoving number density selected sample of galaxies at high redshift, \citet{2011MNRAS.412.1123P} found the average star formation history between $3 < z < 8$ to be best-fitted by a power-law with $\alpha = 1.7 \pm 0.2$ or an exponentially rising history with $\tau = 420$ Myr. Recently, for the deep observations of the CANDELS GOODS South field, \citet{Salmon:2014tm} applied an improved version of this method (incorporating the predicted effects of merger rates on the comoving sample) and found a shallower power-law with $\alpha = 1.4 \pm 0.1$ produced the closest match.

\begin{figure}
    \centering
  \includegraphics[width=0.48\textwidth]{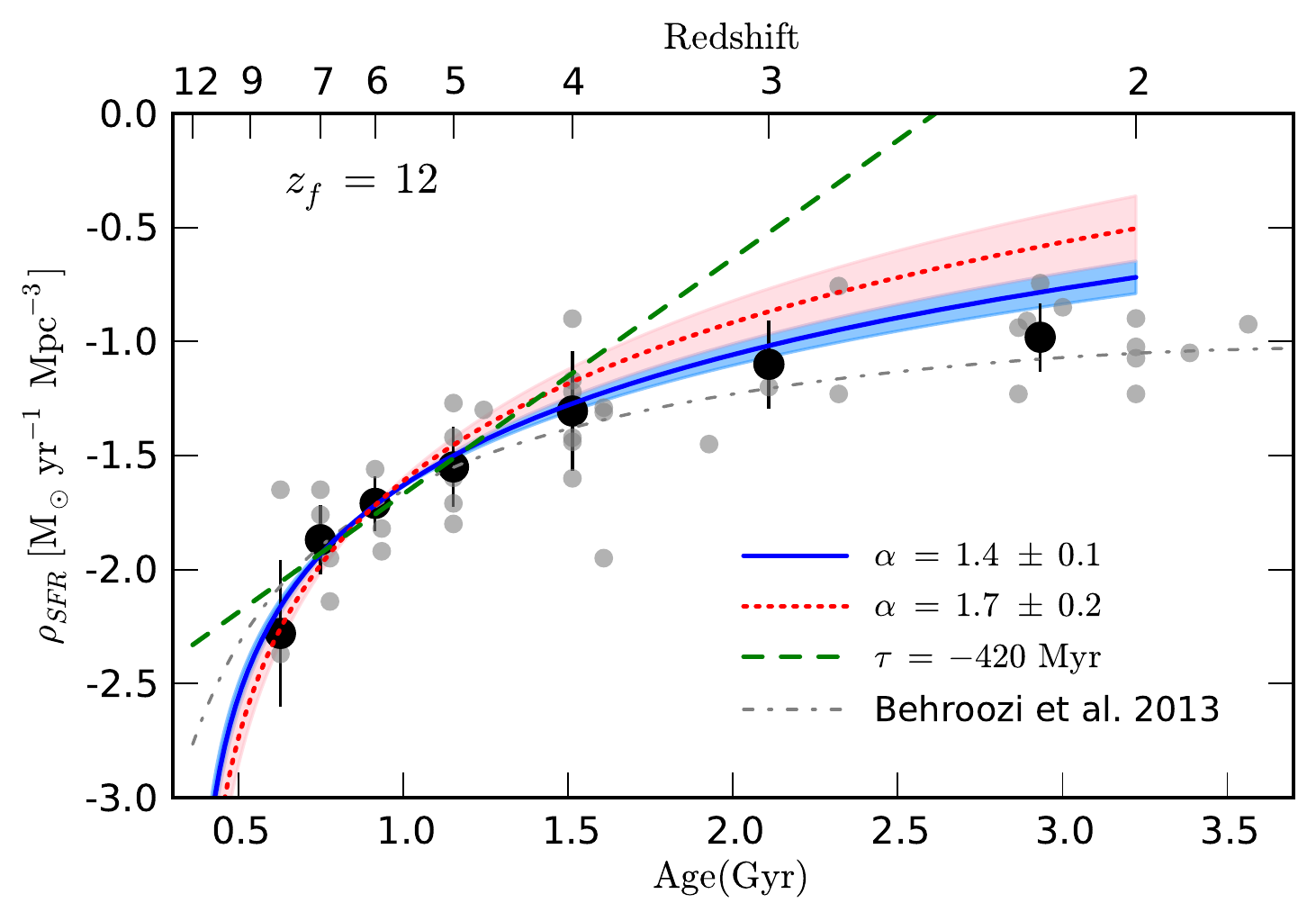}
  \caption{Power-law (blue continuous: \citeauthor{Salmon:2014tm}~\citeyear{Salmon:2014tm}, red dotted: \citeauthor{2011MNRAS.412.1123P}~\citeyear{2011MNRAS.412.1123P}) and exponential (green dashed: \citeauthor{2011MNRAS.412.1123P}~\citeyear{2011MNRAS.412.1123P}) fits to the median observed SFR-densities at $z > 4$ for 3 different star formation histories. The grey datapoints are taken from the compilation of SFR-density observations in \citet{Behroozi:2013fg} with additional points from recent observations \citep{Smit:2012is,Bouwens:2014tx,Duncan:2014gh}. For the power-law fits, the shaded red and blue regions correspond to the 1-$\sigma$ errors on the slope of the power-law, $\alpha$, quoted in the respective papers.  The grey dot-dashed line shows the best-fit to the median star-formation rates across the the full cosmic history for the functional form outlined in \citet{Behroozi:2013fg}. All of the models assume an initial onset of star-formation at $z_{f} = 12$.}
  \label{fig:SFH}
\end{figure}

In Fig.~\ref{fig:SFH}, we show that all three of these models ($\alpha \/ = \/ 1.4/1.7$ and $\tau = -450 \rm{Myr}$) can provide a good fit to the observed evolution in the cosmic star-formation rate density at $z > 3$ (Age of the Universe $\lesssim 2$ Gyr) through a simple scaling alone. However, the power-law fit with $\alpha = 1.4$ provides the best fit to not only to the evolution of the SFR-density at ages $< 2$ Gyr, but also to the SFR density at later epochs. A smoothly rising star-formation is also favoured by hydrodynamic models such as \citet{2011MNRAS.410.1703F} and \citet{Dayal:2013jm}, although the star-formation histories of individual galaxies are likely to be more varied or stochastic \citep{Dayal:2013jm,Kimm:2014gv}. Furthermore, there is also growing evidence of galaxy populations with older populations and possibly quiescent populations \citep{Nayyeri:2014cz,Spitler:2014ey} suggesting some galaxies form very rapidly at high-redshift before becoming quenched. The assumption of a single parametrised SFH is clearly not ideal, however our choice of a rising power-law SFH with $\alpha = 1.4$ is at least well motivated by observations and a more physical choice than a constant or exponentially declining SFH.
    
\subsubsection{Initial mass function}
Interpretation of extragalactic observations through modelling and SED fitting is typically done assuming a universal bi-modal Milky Way-like initial mass function (IMF) such as \citet{Kroupa:2001ki}/\citet{Chabrier:2003ki} or the unimodal \citet{Salpeter:1955hz} IMF. However, there is now growing evidence for systematic variation in the IMF of both nearby \citep{vanDokkum:2010ha,Treu:2010kf,Cappellari:2012jm,Conroy:2012bn,Ferreras:2013id} and distant \citep{MartinNavarro:2014wr} early-type galaxies. 

Under a hierarchical model of galaxy evolution with downsizing, the bright galaxies in overdense regions observed at $z > 3$ are likely to eventually form into the massive early-type galaxies in which these variations can be found. Variation in the slope of the IMF would have a significant effect on many of the critical observable properties at high redshift such as stellar masses and mass to UV light ratio's. However, given the lack of theoretical understanding as to how the IMF should vary with physical conditions, incorporating the effects of a varying IMF at high-redshift is beyond the scope of this work. Throughout the following analysis we use the \citet{Chabrier:2003ki} IMF as our primary assumption, but also consider the systematic effect of a steeper IMF such as \citet{Salpeter:1955hz} on the inferred values or observables in Appendix~\ref{app:tables}.
    
\subsubsection{Metallicity}\label{sec:metallicity}
Current spectroscopic constrains on galaxy metallicities at $z \geq 3$ indicate moderately sub-solar stellar and gas-phase metallicities \citep{Shapley:2003wi,Maiolino:2008gs,Laskar:2011cc,Jones:2012kn,2012A&A...539A.136S}. In addition, \citet{Troncoso:2014kg}, present measurements for 40 galaxies at $3 < z < 5$ for which the observed metallicities are consistent with a downward evolution in the mass-metallicity relation (with increasing redshift).
    
Measurements of galaxy metallicities at higher redshift ($z>6$) are even fewer due to the lack of high-resolution rest-frame optical spectroscopy normally required to constrain metallicity. However, thanks to a clear detection of the {\sc Ciii]} emission line ($1909\rm{\AA}$) and strong photometric constraints, \citet{Stark:2014vc} are able to measure a metallicity of $\approx 1/20$th solar metallicity for a lensed galaxy at $z = 6.029$. Given these observations and the metallicities available in the \citet{Bruzual:2003ckb} models, we assume a fiducial metallicity of $\approx 1/5$th solar metallicity ($Z = 0.004 = 0.2 Z_{\odot}$).
        
\subsubsection{Age}
For the rising star-formation history used throughout this work, there is a weak evolution of $\beta$ as a function of age ($\Delta\beta \approx 0.13$ between $t \approx 100$ Myr and $\approx 1$ Gyr) whereby older stellar populations have redder UV continuum slopes. However, at very young ages, the contribution of nebular continuum emission in the UV continuum can also significantly redden the apparent $\beta$ compared to the much steeper underlying intrinsic UV slope \citep{2010Natur.468...49R}.  This results in a degeneracy with respect to $\beta$ between young and old populations. For example, for identical observed (stellar + nebular continuum) UV-slopes, $\xi_{\rm{ion}}$ (and $\kappa_{\rm{ion}}$) for a young stellar population can be a factor of up $\sim 0.5~ (0.2)~\rm{dex}$ higher. Given this degeneracy, additional constraints from other parts of the electromagnetic spectrum are required in order to make a well informed choice of stellar population age.

Due to the observational restrictions on high-resolution rest-frame optical spectroscopy, measurement of stellar population ages for high-z galaxies is limited to photometric fitting and colour analysis. At $z\sim4$, where the Balmer break can be constrained through deep \emph{Spitzer} IRAC photometry, estimates of the average stellar population ages vary significantly from $\sim 200-400$ Myr \citep{Lee:2011dw} to $\sim1$ Gyr \citep{Oesch:2013eb} (dependent on assumptions of star-formation history). 
 
    
For galaxies closer to the epoch of reionization, constraints on the Balmer/D(4000) breaks become poorer due to the fewer bands available to measure the continuum above the break, a problem which is exacerbated by the additional degeneracy of strong nebular emission lines redshifted into those filters \citep{2009A&A...502..423S,2010A&A...515A..73S}. The effect of incorporating the effects of emission lines on SED fits at $z \geq 5$ is that on average the best-fitting ages and stellar masses are lowered. This is because the rest-frame optical colours can often be well fit by either a strong Balmer break or by a significantly younger stellar population with very high equivalent width H$\alpha$ or {\sc Oiii} emission. 


Recent observations of galaxies at high-redshift with constraints on the UV emission-line strengths (Ly$\alpha$ or otherwise) have found that single-component star-formation histories are unable to adequately fit both the strong line emission and the photometry at longer wavelengths \citep{RodriguezEspinosa:2014cs,Stark:2014vc}. For example, in order to match both the observed photometry at rest-frame optical wavelengths and the high-equivalent width UV emission lines of a lensed galaxy at $z = 6.02$, \citet{Stark:2014vc} require two stellar populations. In combination with a `young' 10 Myr old starburst component, the older stellar component is best fitted with an age since the onset of star-formation of $\approx$ 500 Myr.

Based on the observations discussed above and the median best-fit ages found by \citet{2013MNRAS.429..302C}, the star-formation history outlined in Section~\ref{sec:sfh} is consistent with the existing limited constraints. At $z\sim7$, the redshift of strongest interest to current studies of reionization, the assumed onset of star-formation at $z=12$ gives rise to a stellar population age (since the onset of star-formation) of $\sim390$ Myr. 

\subsubsection{Nebular continuum and line emission}\label{sec:neb}
If nebular line emission is ``ubiquitous" at high-redshift as an increasing number of studies claim (e.g. \citet{Shim:2011cw,Stark:2013ix,Smit:2013ud}), the accompanying nebular continuum emission should also have a strong effect on the observed SEDs of high-redshift galaxies \citep{Reines:2009gs}. In this work, we include both nebular continuum and optical line emission using the prescription outlined in \citet{Duncan:2014gh} (and equivalent to the methods described in \citet{Ono:2010ed,2010A&A...515A..73S,2011MNRAS.418.2074M}). 

Whilst the strength of nebular emission in this model is directly proportional to the number of ionizing photons produced by the underlying stellar population, additionally both the strength and spectral shape of the nebular continuum emission are also dependent on the continuum emission coefficient, $\gamma^{(total)}_{\nu}$, given by
\begin{equation}\label{eq:cont_sep}
\gamma^{(total)}_{\nu} = \gamma^{(HI)}_{\nu} + \gamma^{(2q)}_{\nu} +  \gamma^{(HeI)}_{\nu}\frac{n(He^{+})} {n(H^{+})} + \gamma^{(HeII)}_{\nu}\frac{n(He^{++})} {n(H^{+})}
\end{equation}
where $\gamma^{(HI)}_{\nu}$, $\gamma^{(HeI)}_{\nu}$, $\gamma^{(HeII)}_{\nu}$ and $\gamma^{(2q)}_{\nu}$ are the continuum emission coefficients for free-free and free-bound emission by Hydrogen, neutral Helium, singly ionized Helium and two-photon emission for Hydrogen respectively \citep{1995A&A...303...41K}. As in \citet{Duncan:2014gh}, the assumed continuum coefficients are taken from \citet{Osterbrock:2006ula}, assuming an electron temperature $T=10^4$ K and electron density $n_{e}=10^2$ cm$^{-3}$ and abundance ratios of $y^{+} \equiv \frac{n(He^{+})} {n(H^{+})} = 0.1$ and $y^{++} \equiv \frac{n(He^{++})} {n(H^{+})} = 0$ \citep{1995A&A...303...41K,Ono:2010ed}. 

Although the exact ISM conditions and abundances of high-redshift {\sc Hii} regions is not well known, singly and doubly ionised helium abundances for nearby low-metallicity galaxies have been found to be $y^{+} \approx 0.08$ and $y^{++} \approx 0.001$ \citep{Dinerstein:1986di,Izotov:1994fi,Hagele:2006eq}. We estimate that for the age, metallicity and dust values chosen for our fiducial model (see Table~\ref{tab:fiducial}), variations of $\Delta y^{+} = 0.05$ corresponds to $\Delta\beta = 0.004$, while values of $y^{++}$ as large as $3\%$ \citep{Izotov:2013ce} would redden the observed UV slope by $\Delta\beta = +0.003$. In this case, because the nebular continuum emission is dominated by the stellar continuum at these wavelengths for our assumption, the effects of variation in the {\sc Hii} region properties is negligible and our interpretation of the observed UV slopes should not be affected by our assumed nebular emission properties.

\subsubsection{Dust Extinction}\label{sec:models_dust}
In \citet{Bouwens:2009ik}, it is argued that the most likely physical explanation for variation in the observed $\beta$s between galaxies is through the variation in dust content. Large changes in metallicity and ages are required to produce significant variation in $\beta$ (see later discussion in Section~\ref{sec:slp}), however as previously discussed in this section, such large variations are not observed in the galaxy population in either age or metallicity at $z > 3$ based on current observations. For the fiducial model in our subsequent analysis, we allow the magnitude of dust extinction ($A_{V}$) to vary along with $f_{\rm{esc}}$, but we must also choose a dust attenuation law to apply.

Direct measurements of dust and gas at extreme redshifts are now possible thanks to the sub-mm facilities of ALMA. However, the current number of high redshift observations is still very small. \citet{2013ApJ...778..102O} and \citet{Ota:2014du} observe Lyman alpha emitters (LAEs) at $z\sim7$, finding only modest dust extinction ($E(B-V) = 0.15$). Recent work by \citet{Schaerer:2014wo} extends the analysis to a larger sample of five galaxies, finding a range in dust extinction of $0.1 < A_{V} < 0.8$. While the dust attenuation in these objects is consistent with normal extragalactic attenuation curves such as \citep{2000ApJ...533..682C} or the SMC extinction curve (e.g. \citet{Pei:1992ey}), the results are not strong enough to constrain or distinguish between these models. Similarly, for broadband SED fits of high-redshift objects, neither a starburst or SMC-like attenuation curve is strongly favoured \citep{Salmon:2014tm}. Based on these factors, we assume the starburst dust attenuation curve of \citet{2000ApJ...533..682C} in order to make consistent comparisons with the SED fitting of \citet{Duncan:2014gh} and \citet{Meurer:1999jm} dust corrections to UV star-formation rates (e.g. \citet{Bouwens:2011tj,Smit:2012is}). 

In addition, due to the lack of constraints on the dust attenuation strengths at wavelengths less than $1200\rm{\AA}$ we must also assume a plausible extrapolation below these wavelengths. For our fiducial model, we simply extrapolate linearly based on the slope of the attenuation curve at $1200-1250 \rm{\AA}$, in line with similar works on the escape fraction of galaxies \citep{Siana:2007bc}. In addition, we also assume a second model in which the extreme-UV and Lyman continuum extinction follows the functional form of the component of the \citet{Pei:1992ey} SMC extinction model at $\lambda \leq 1000 \rm{\AA}$, whereby the relative absorption begins to decrease below $800\rm{\AA}$. The systematic effect of choosing this second assumption along with a third assumption of an SMC extinction curve are outlined in Table~\ref{tab:bouwens_N}.

As shown in Fig.~\ref{fig:mechanisms}, we assume a simple foreground dust screen \citep{1994ApJ...429..582C} and that dust destruction is minimal and/or balanced by grain production \citep{Zafar:2013fe,Rowlands:2014dq}; effectively that dust for a given model is fixed with relation to the stellar population age. The assumption of a different dust geometry, such as one with clouds dispersed throughout the ISM, would require a greater amount of dust to achieve the same optical depth and could also have a significant effect on the extinction of nebular emission relative to that of the stellar continuum \citep{Zackrisson:2013iz}.

\subsubsection{Differing SSP models: the effects of stellar rotation and binarity}
The choice of \citet{Bruzual:2003ckb} stellar population synthesis (SPS) models in this work was motivated by the more direct comparison which can be made between this analysis and the stellar mass, luminosity and colour measurements based on the same models, e.g. \citet{2012ApJ...756..164F,Duncan:2014gh}. However, several other SPS models are available and in common usage, e.g. Starburst99 \citep{Leitherer:1999jt}, \citet{Maraston:2005er} and FSPS \citep{Conroy:2009ks,Conroy:2009ja}.

Due to differences in assumptions/treatment of various ingredients such as horizontal branch morphology or thermally-pulsating asymptotic giant branch (TP-AGB) stars, the SEDs produced for the same input galaxy properties (such as age and metallicity) can vary significantly. The full systematic effects of the different assumptions and models for galaxies at high-redshift is not well quantified and adequately doing so is beyond the scope of this work. We do however caution that these systematics could significantly affect the inferred ionizing photons rates of galaxies during the EoR. In particular, it has been suggested that rotation of massive stars could have a significant effect on the UV spectra and production rate of ionizing photons \citep{Vazquez:2007jt}.

In \citet{Leitherer:2014ia}, the effects of the new stellar models including stellar rotation \citep{Ekstrom:2012ke} are incorporated into the Starburst99 SPS models. The resulting SEDs are changed drastically with an increase in the ionizing photon rate of up to a factor of five for the most extreme model of rotation.

A second, equally significant effect comes from the inclusion of binary physics in stellar population synthesis models. It is now believed that the majority of massive stars exist in binaries \citep{Sana:2012gu,Sana:2013hh,Aldoretta:2014un} while the majority of stellar population models (including all of the aforementioned SPS models) are for single stars. The {\sc bpass} code of \citet{Eldridge:2009bi,Eldridge:2011kg}) incorporates the physics of binary rotation on massive stars to explore the effects on the predicted stellar population features, observing similar effects to the addition of rotation in single star models; an increased fraction of red supergiants which go on to form bluer UV bright Wolf-Rayet stars.

We note that for the same assumed stellar populations parameters (age, metallicity, SFH, dust), the use of either of these models would result in SEDs with bluer UV continuum slopes and an increased LyC production rate. However, the full ramifications of how these models may change the interpretation of SEDs, stellar masses and $\beta$s for the observations of high-redshift galaxies is beyond the scope of this work. 

\subsection{Observed UV slopes as a function of $f_{\rm{esc}}$ and dust extinction}\label{sec:slp}
\begin{figure*}
\centering
  \includegraphics[width=0.33\textwidth]{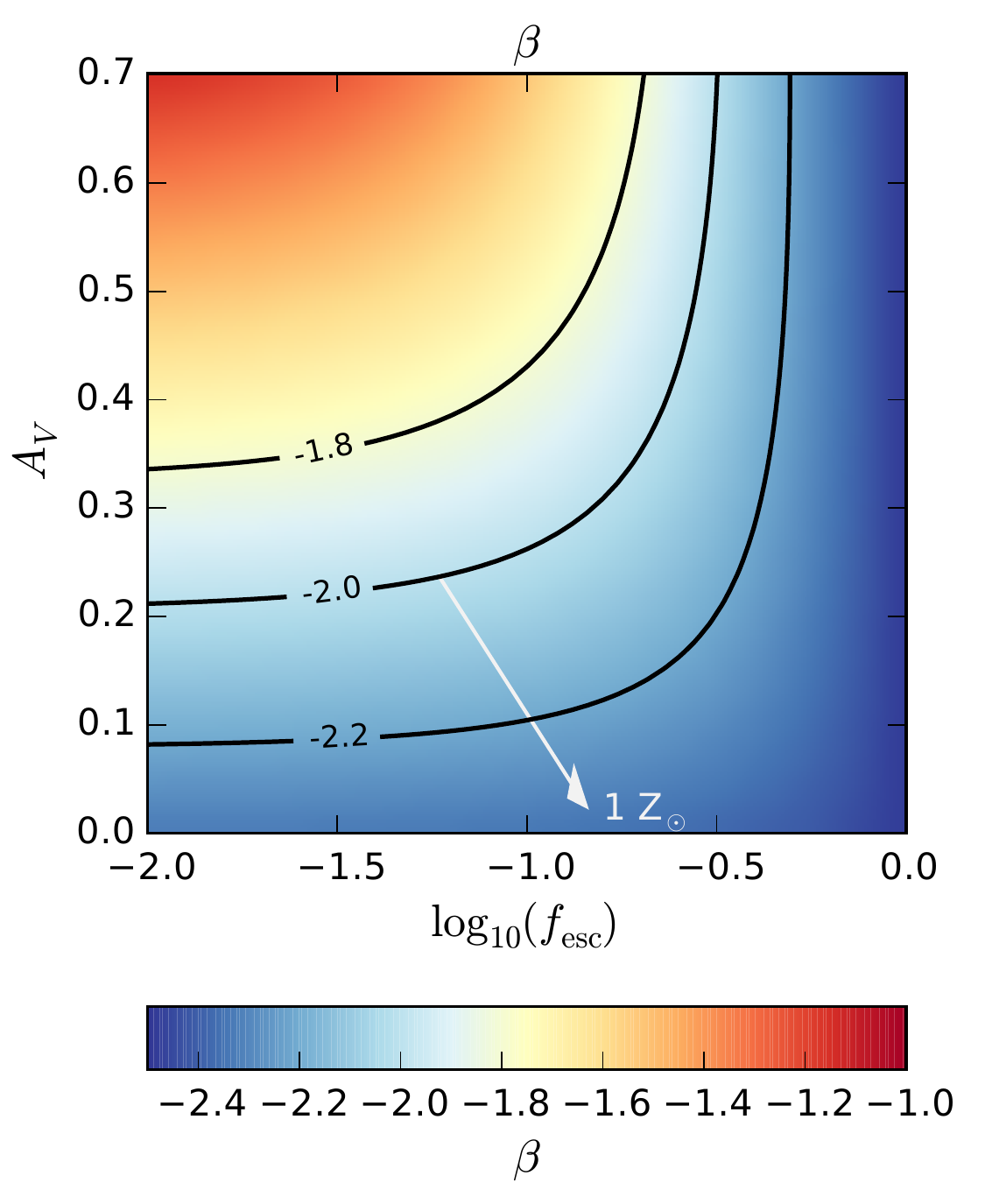}
  \includegraphics[width=0.33\textwidth]{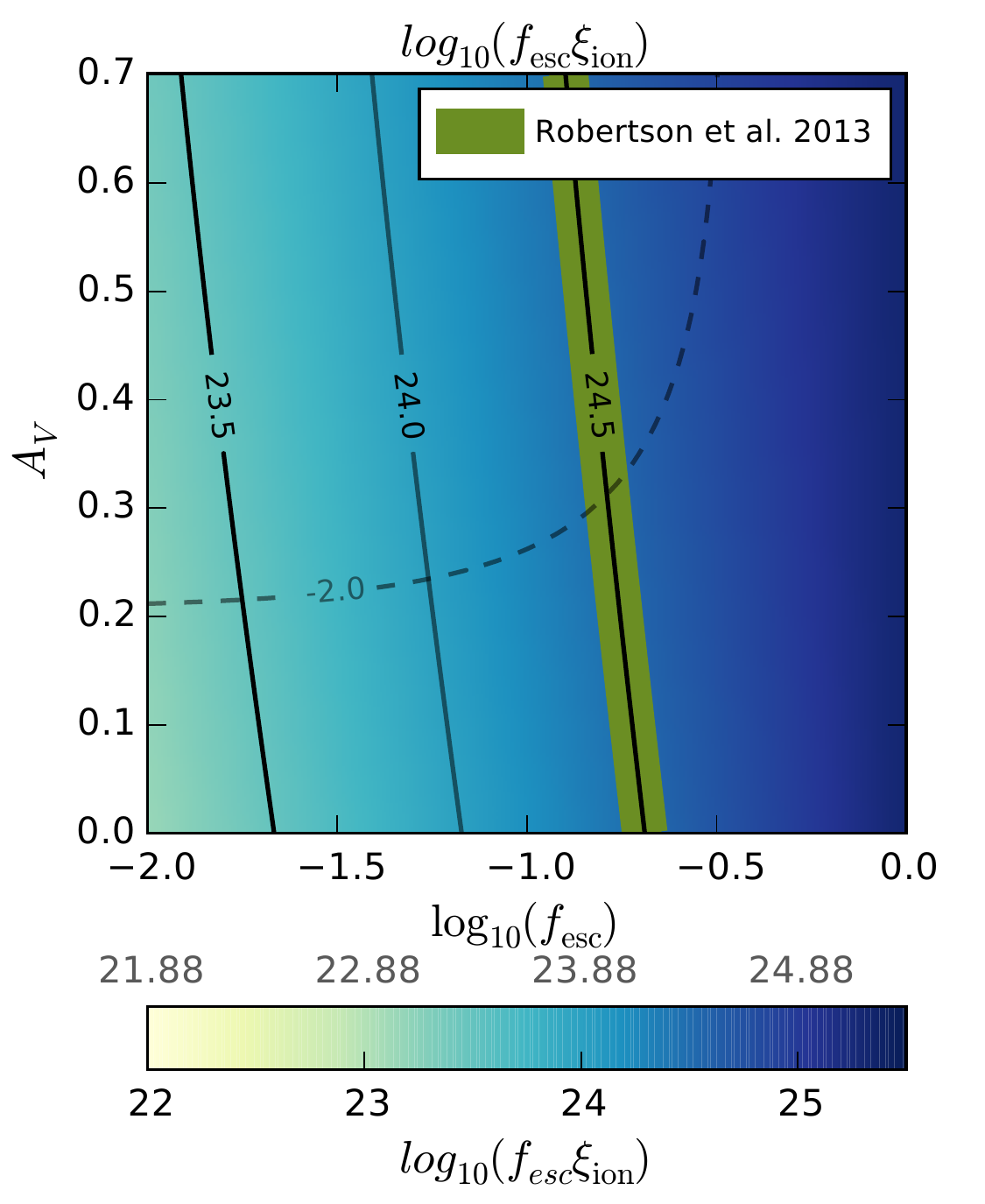}
  \includegraphics[width=0.33\textwidth]{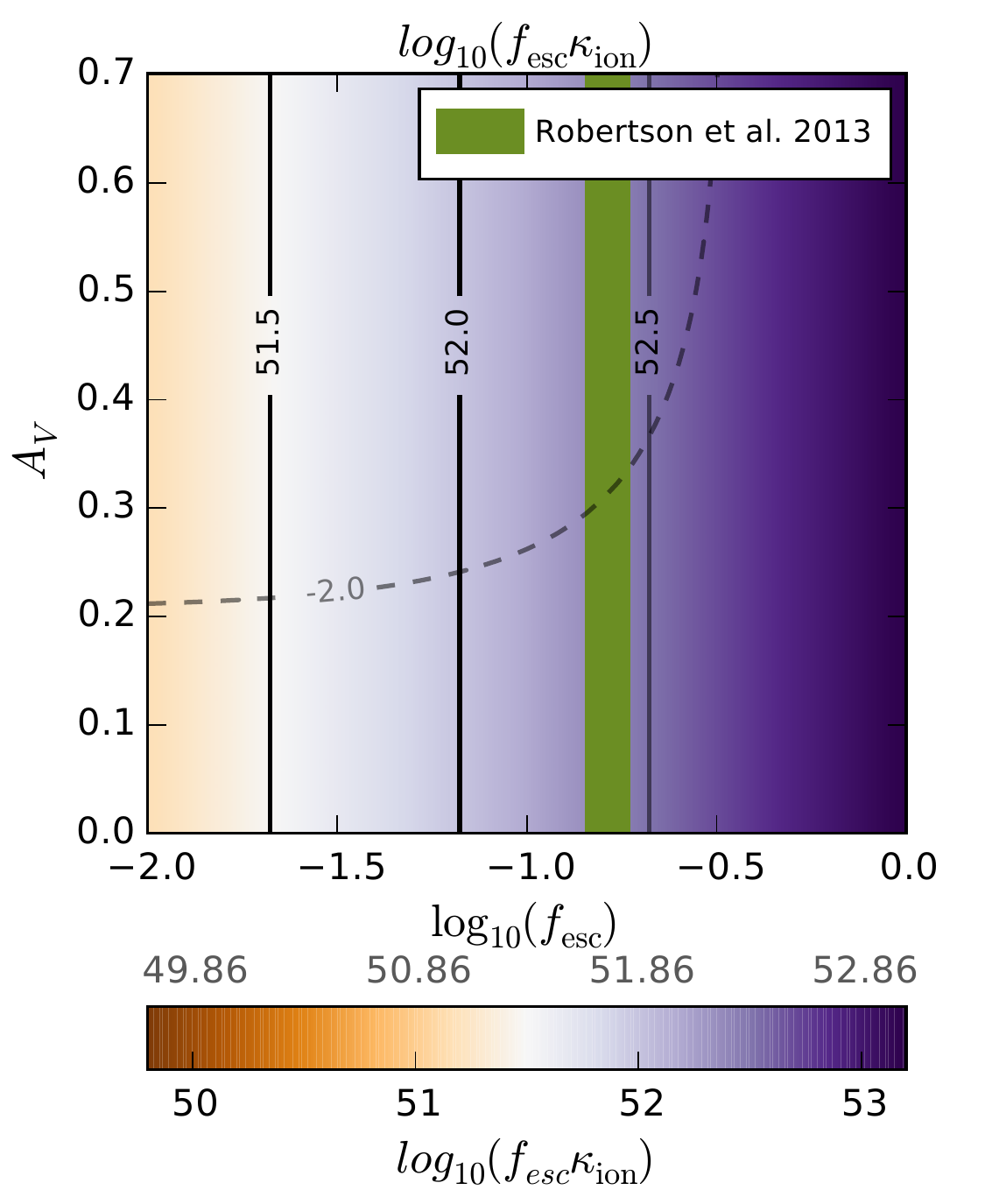}

  \caption{Left: UV continuum slope $\beta$ as a function of total escape fraction, $f_{\rm{esc}}$, and dust extinction, $A_{V}$, for the \emph{ionization bounded nebula with holes} continuum escape model (Model A, Fig.~\ref{fig:mechanisms} left) with stellar population properties as outlined in Section~\ref{sec:assumptions}. The contours indicate lines of constant $\beta$ around the observed average $\beta$, and the light grey arrow indicates how those contours move for a stellar population with solar metallicity. Middle and right: $\log_{10}f_{\rm{esc}}\xi_{\rm{ion}}$ and $\log_{10}f_{\rm{esc}}\kappa_{\rm{ion}}$ as a function of escape fraction and dust extinction respectively for the same continuum escape model. Solid contours represent lines of constant $f_{\rm{esc}}\xi_{\rm{ion}}$ and $f_{\rm{esc}}\kappa_{\rm{ion}}$ whilst the dashed contour shows where $\beta = -2$ is located for reference. The green labelled contour shows the assumed $f_{\rm{esc}}\xi_{\rm{ion}} = 24.5$ value of \citetalias{Robertson:2013ji} and the equivalent in $f_{\rm{esc}}\kappa_{\rm{ion}}$ (see text for details). For the colour scales below the centre and right panels, the lower black tick labels correspond to the scale for the fiducial model ($Z = 0.2 Z_{\odot}$) whilst the grey upper tick label indicate how $f_{\rm{esc}}\xi_{\rm{ion}}$ and $f_{\rm{esc}}\kappa_{\rm{ion}}$ change for stellar populations with solar metallicity.}
  \label{fig:fesc_cover}
\end{figure*}

\begin{figure*}
\centering
  \includegraphics[width=0.33\textwidth]{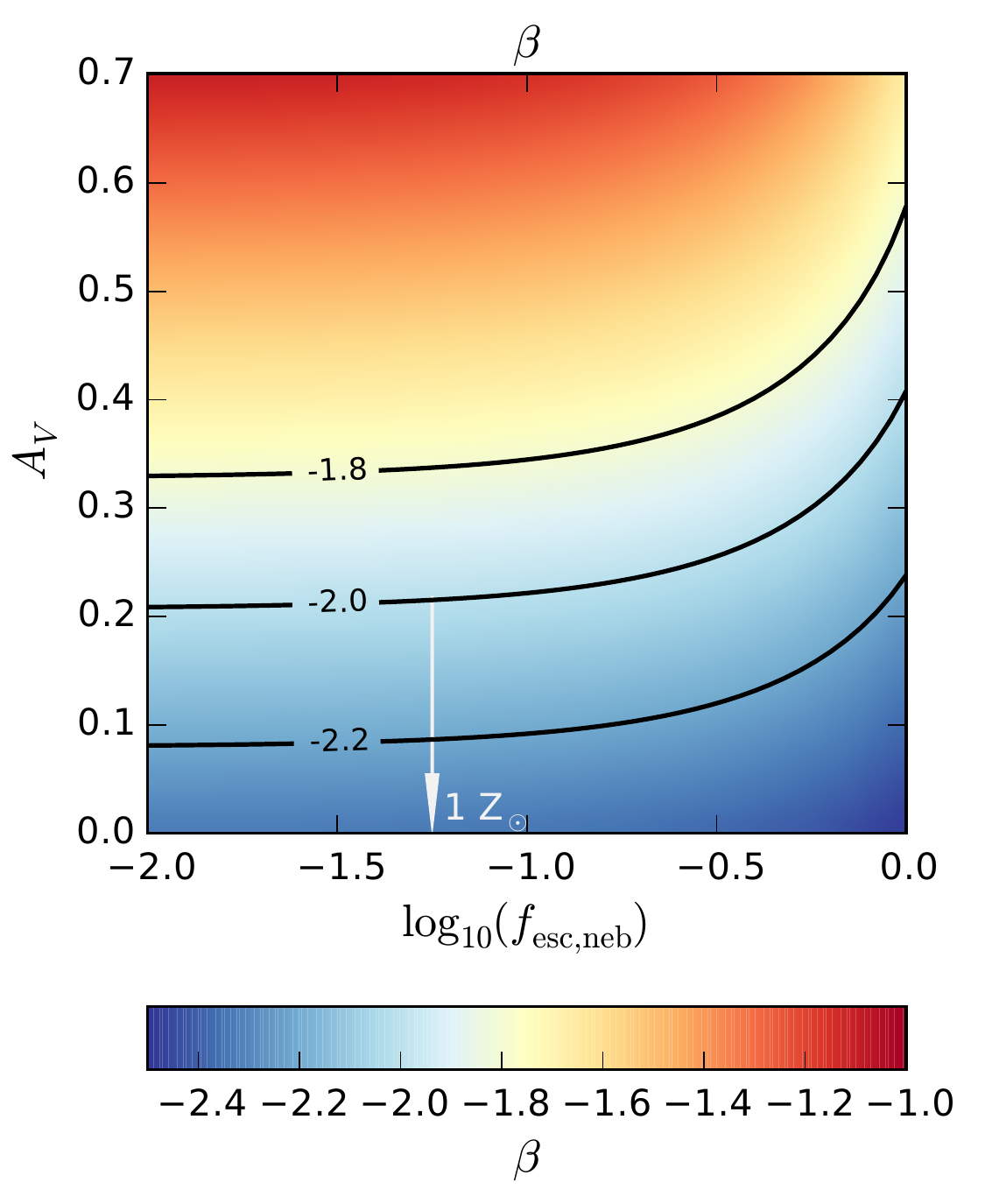}
  \includegraphics[width=0.33\textwidth]{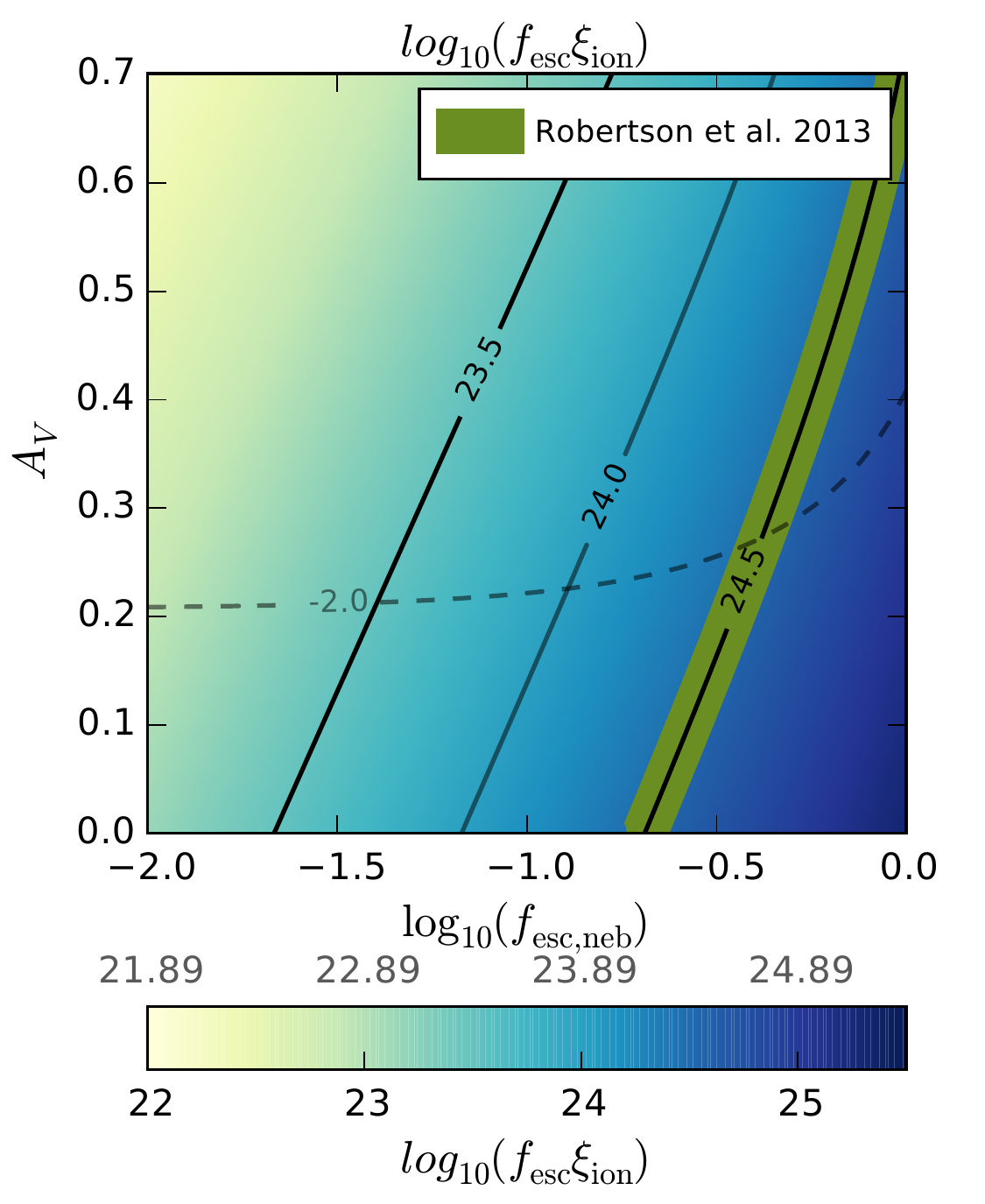}
  \includegraphics[width=0.33\textwidth]{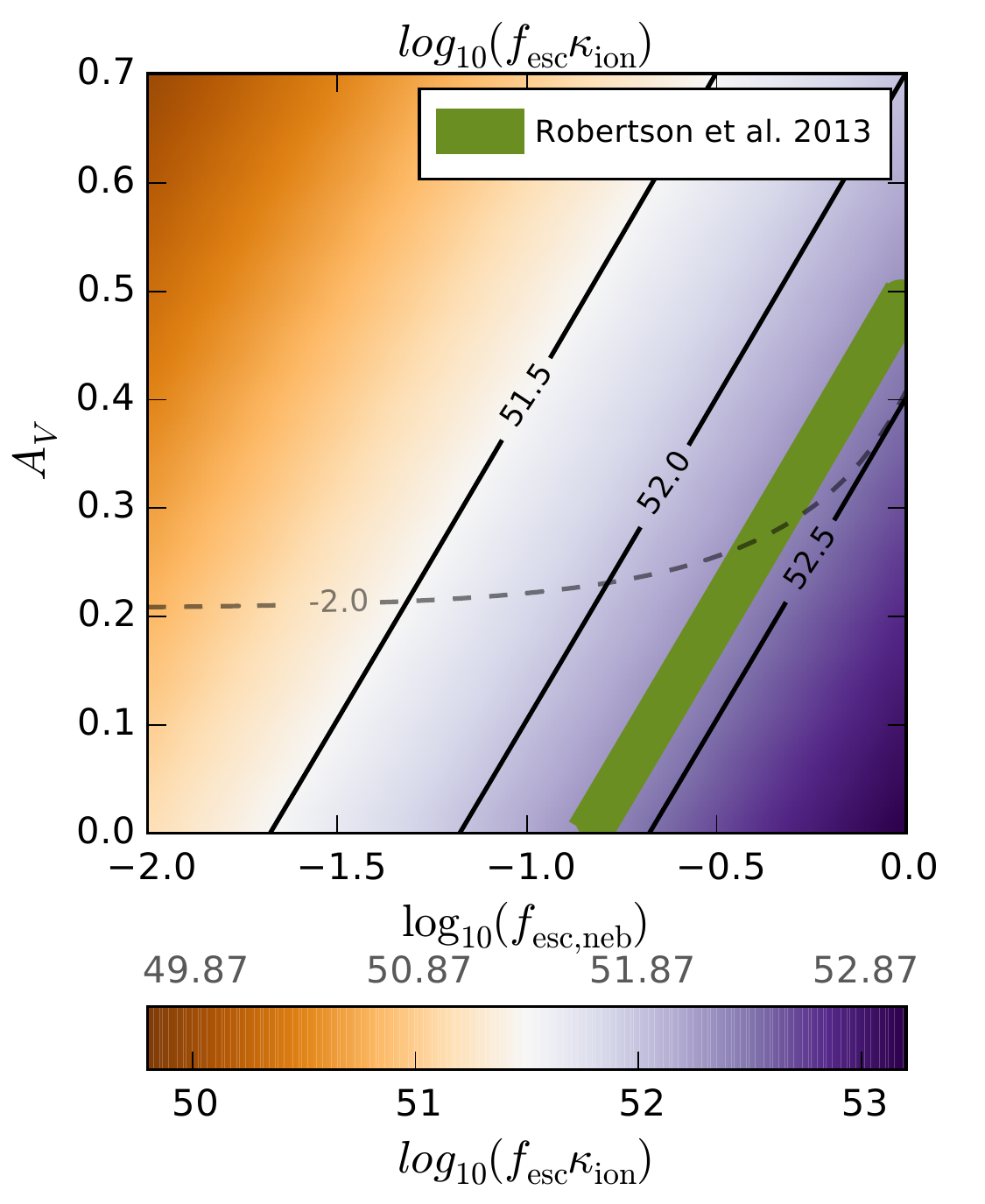}

  \caption{Left: UV continuum slope $\beta$ as a function of {\sc Hii} region escape fraction, $f_{\rm{esc}}$, and dust extinction, $A_{V}$, for the \emph{density bounded nebula} continuum escape model (Model B, Fig.~\ref{fig:mechanisms} right) with stellar population properties as outlined in Section~\ref{sec:assumptions}. The contours indicate lines of constant $\beta$ around the observed average $\beta$, and the light grey arrow indicates how those contours move for a stellar population with solar metallicity. Centre and right: $\log_{10}f_{\rm{esc}}\xi_{\rm{ion}}$ and $\log_{10}f_{\rm{esc}}\kappa_{\rm{ion}}$ (where $f_{\rm{esc}}$ is the total dust attenuated escape fraction) as a function of escape fraction and dust extinction respectively for the same continuum escape model. Solid contours represent lines of constant $f_{\rm{esc}}\xi_{\rm{ion}}$ and $f_{\rm{esc}}\kappa_{\rm{ion}}$ whilst the dashed contour shows where $\beta = -2$ is located for reference. The green labelled contour shows the assumed $f_{\rm{esc}}\xi_{\rm{ion}} = 24.5$ value of \citetalias{Robertson:2013ji} and the equivalent in $f_{\rm{esc}}\kappa_{\rm{ion}}$ (see text for details). For the colour scales below the centre and right panels, the lower black tick labels correspond to the scale for the fiducial model ($Z = 0.2 Z_{\odot}$) whilst the grey upper tick label indicate how $f_{\rm{esc}}\xi_{\rm{ion}}$ and $f_{\rm{esc}}\kappa_{\rm{ion}}$ change for stellar populations with solar metallicity.}
  \label{fig:fesc_dens}
\end{figure*}

The galaxy properties with the largest uncertainties and expected variation are the optical depth of the dust attenuation (or extinction magnitude $A_{V}$)  and the property we wish to constrain photometrically, the escape fraction of ionizing photons $f_{\rm{esc}}$. For the assumptions of our fiducial model (Table~\ref{tab:fiducial}), we want to explore the possible range of these two properties which are consistent with the observed UV slopes (as calculated for each model following Eq.~\ref{eq:beta_calc}) and what constraints can then be placed on the ionizing emissivity coefficients $f_{\rm{esc}}\xi_{\rm{ion}}$ or $f_{\rm{esc}}\kappa_{\rm{ion}}$.

In the left panels of Fig.~\ref{fig:fesc_cover} and Fig.~\ref{fig:fesc_dens}, we show how $\beta$ varies as a function of the dust extinction magnitude and escape fraction for each of the continuum escape mechanisms respectively (Section~\ref{sec:link}/Fig.~\ref{fig:mechanisms}). For both models, $\beta$ is relatively constant as a function of $f_{\rm{esc}}$ at low values of escape fraction ($f_{\rm{esc}} < 0.1$) . At larger escape fractions, the two mechanisms produce different UV slopes. For Model A, as $f_{\rm{esc}}$ increases to $\sim10\%$ (covering fraction $\approx 90\%$) the unattenuated stellar continuum begins to dominate the overall colours as $f_{\rm{esc}}$ increases and by $f_{\rm{esc}} \gtrsim 30\%$ the observed average $\beta$ is determined only by the unattenuated light, irrespective of the magnitude of the dust extinction in the covered fraction. This effect is also illustrated in a different way in Figure 9 of \citet{Zackrisson:2013iz}, whereby the same amount of dust extinction in the covered/high-density regions has a decreasing effect on the observed $\beta$ as the escape fraction increases. This means that if Lyman continuum is escaping through holes in the ISM and is un-attenuated by dust, it is possible set constraints on the maximum escape fraction possible which is still consistent with the UV slopes observed.

For Model B, $\beta$ remains constant with $f_{esc,neb}$ at a fixed dust extinction until $f_{\rm{esc}} \approx 30\%$. Beyond this, the reduction in nebular continuum emission from the high escape fraction begins to make the observed $\beta$s bluer for the same magnitude of dust extinction.

For both escape mechanisms, a UV slope of $\beta \approx -2$ is achievable with only moderate amounts of dust extinction required ($A_{V} \approx 0.4$ and 0.25 for models A and B respectively at $f_{\rm{esc}}\approx 0.2$). When metallicity is increased to $Z = Z_{\odot}$, the isochromes (of constant $\beta$) are shifted downwards such that $\beta \approx -2$ requires negligible dust attenuation (cf. \citeauthor{Robertson:2013ji}~\citeyear{Robertson:2013ji}). In Section~\ref{sec:beta_effect_stellar_pop} we further explore the effects of varying metallicity on the apparent $\beta$ and corresponding emissivity coefficients. However, first we wish to examine the range of $\xi_{\lowercase{ion}}$ and $\kappa_{\lowercase{ion}}$ which correspond to the values of $f_{\rm{esc}}$ and $A_{V}$ consistent with $\beta \approx -2$ found here.

\subsection{$\xi_{\lowercase{ion}}$ and $\kappa_{\lowercase{ion}}$ as a function of $f_{\rm{esc}}$ and dust extinction}
In the centre and right panels of Figures~\ref{fig:fesc_cover} and \ref{fig:fesc_dens} we show $\log_{10}f_{\rm{esc}}\xi_{\rm{ion}}$ and $\log_{10}f_{\rm{esc}}\kappa_{\rm{ion}}$ as a function of $f_{\rm{esc}}$ and the extinction magnitude of dust in the covered regions ($A_{V}$).

For dust model A, the ionization bounded nebula with holes, there is very little dependence of the ionizing photon rate per unit UV luminosity on the magnitude of dust extinction in the covered fraction (centre panel). Because the increase dust extinction magnitude around the high column density areas only affects the UV/optical component, the increasing dust extinction results in higher values of $\log_{10}f_{\rm{esc}}\xi_{\rm{ion}}$ due to the increased absorption of the UV light in the dust covered regions. For this dust model, the corresponding ionizing photon rate per unit SFR ($f_{\rm{esc}}\kappa_{\rm{ion}}$) has zero evolution as a function of dust extinction in this geometry.

The assumed  $\log_{10}f_{\rm{esc}}\xi_{\rm{ion}} = 24.5$ of \citetalias{Robertson:2013ji} is consistent with  $\beta  = -2$ for this model, with an escape fraction of $f_{\rm{esc}} = 0.16$ and moderate dust extinction ($A_{V} = 0.31$). Given the low escape fractions which are still consistent with blue $\beta$ slopes, a value of $\log_{10}f_{\rm{esc}}\xi_{\rm{ion}} = 24.5$ does represent a relatively optimistic assumption on the ionizing efficiency of galaxies. However, it is still $\approx 0.2$ dex lower than the largest $f_{\rm{esc}}$ still consistent with a UV slope of $\beta = -2$.

In contrast to model A, because the dust in the density bounded nebula (model B) is assumed to cover all angles, increases in the dust extinction magnitude results in significantly smaller $\log_{10}f_{\rm{esc}}\xi_{\rm{ion}}$/$\log_{10}f_{\rm{esc}}\kappa_{\rm{ion}}$ for the same fixed $f_{\rm{esc}}$. This can be seen clearly in the centre and right panels of Fig.~\ref{fig:fesc_dens}.

Despite this, an assumed value of $\log_{10}f_{\rm{esc}}\xi_{\rm{ion}} = 24.5$ is still consistent with $\beta = -2$ for this model. However, it requires a higher escape fraction ($f_{\rm{esc}} = 0.42$) and lower dust extinction ($A_{V} = 0.27$) to achieve this for the same underlying stellar population. The maximum $f_{esc,neb} = 1$ is still consistent with the fiducial UV slope, but the increased dust required to match $\beta = -2$ means that the total LyC escape fraction is reduced and that the corresponding maximum $\log_{10}f_{\rm{esc}}\xi_{\rm{ion}}$ is only marginally higher than the assumptions of \citep{Kuhlen:2012ka} or \citepalias{Robertson:2013ji}. 

\begin{table}
  \caption{Summary of the stellar population assumptions for our fiducial $\beta = -2$ model.}
    \centering
    \begin{minipage}\textwidth
  \begin{tabular}{r|c c}
     Star-formation history & \multicolumn{2}{c}{$SFR \propto t^{1.4}$ \footnote{\citet{Salmon:2014tm}}}\\
    Initial Mass Function & \multicolumn{2}{c}{\citet{Chabrier:2003ki}} \\
    Dust attenuation curve & \multicolumn{2}{c}{\citet{2000ApJ...533..682C}} \\
    Metallicity & \multicolumn{2}{c}{ $Z = 0.2 Z_{\odot}$ }\\
    Nebular Emission & \multicolumn{2}{c}{Continuum included\footnote{$T=10^4$ K, $n_{e}=10^2$ cm$^{-3}$, $y^{+} = 0.1$ and $y^{++} = 0$}}\\
    Age & \multicolumn{2}{c}{390 Myr\footnote{Age at $z\sim7$ since onset of star-formation at $z\sim12$.}}\\
    & & \\
    & \emph{Model A} & \emph{Model B} \\
    Dust attenuation magnitude $A_{V}$ & 0.31 & 0.27 \\
    Escape fraction $f_{esc,neb}$ & 0.16 & 0.42  \\
    $\log_{10} f_{\rm{esc}}\xi_{\rm{ion}}$ & 24.5\footnote{The assumed  $\log_{10} f_{\rm{esc}}\xi_{\rm{ion}}$ of \citetalias{Robertson:2013ji}} & 24.5 \\ 
    $\log_{10} f_{\rm{esc}}\kappa_{\rm{ion}}$ & 52.39 & 52.34 \\    
  \end{tabular}
  \label{tab:fiducial}
  \end{minipage}
\end{table}

For the assumed stellar population properties in our reference model, the UV continuum slope of the intrinsic dust-free stellar population is $\beta = -2.55$ excluding the contribution of nebular continuum emission. This value is significantly bluer than the dust-free $\beta$ assumed by the \citet{Meurer:1999jm} relation commonly used to correct UV star-formation rates for dust absorption. It is however in better agreement with the dust-free $\beta$'s estimated recently for observed galaxies at $z\geq 3$ \citep{Castellano:2014db,deBarros:2014fa} and the theoretical predictions of \citet{Dayal:2012kp}. 

\subsection{Effect of different stellar population properties on $\xi_{\lowercase{ion}}$ and $\kappa_{\lowercase{ion}}$ vs $\beta$}\label{sec:beta_effect_stellar_pop}
Given the strong evidence for both luminosity and redshift dependent $\beta$s, we wish to explore whether evolution in each of the stellar population parameters can account for the observed range of $\beta$s and estimate what effect such evolution would have on the inferred values of $f_{\rm{esc}}\xi_{\rm{ion}}$ and $f_{\rm{esc}}\kappa_{\rm{ion}}$. At $z >6$, the the average $\beta$ for the brightest and faintest galaxies by $\Delta\beta \approx 0.6$ (Fig.~\ref{fig:beta_z}). As a function of redshift, the evolution in $\beta$ is less dramatic, with average slopes (at a fixed luminosity) reddening by $\Delta\beta \approx 0.1$ in the $\sim400$ million years between $z\sim7$ and $z\sim5$.

\begin{figure*}
  \includegraphics[width=0.42\textwidth]{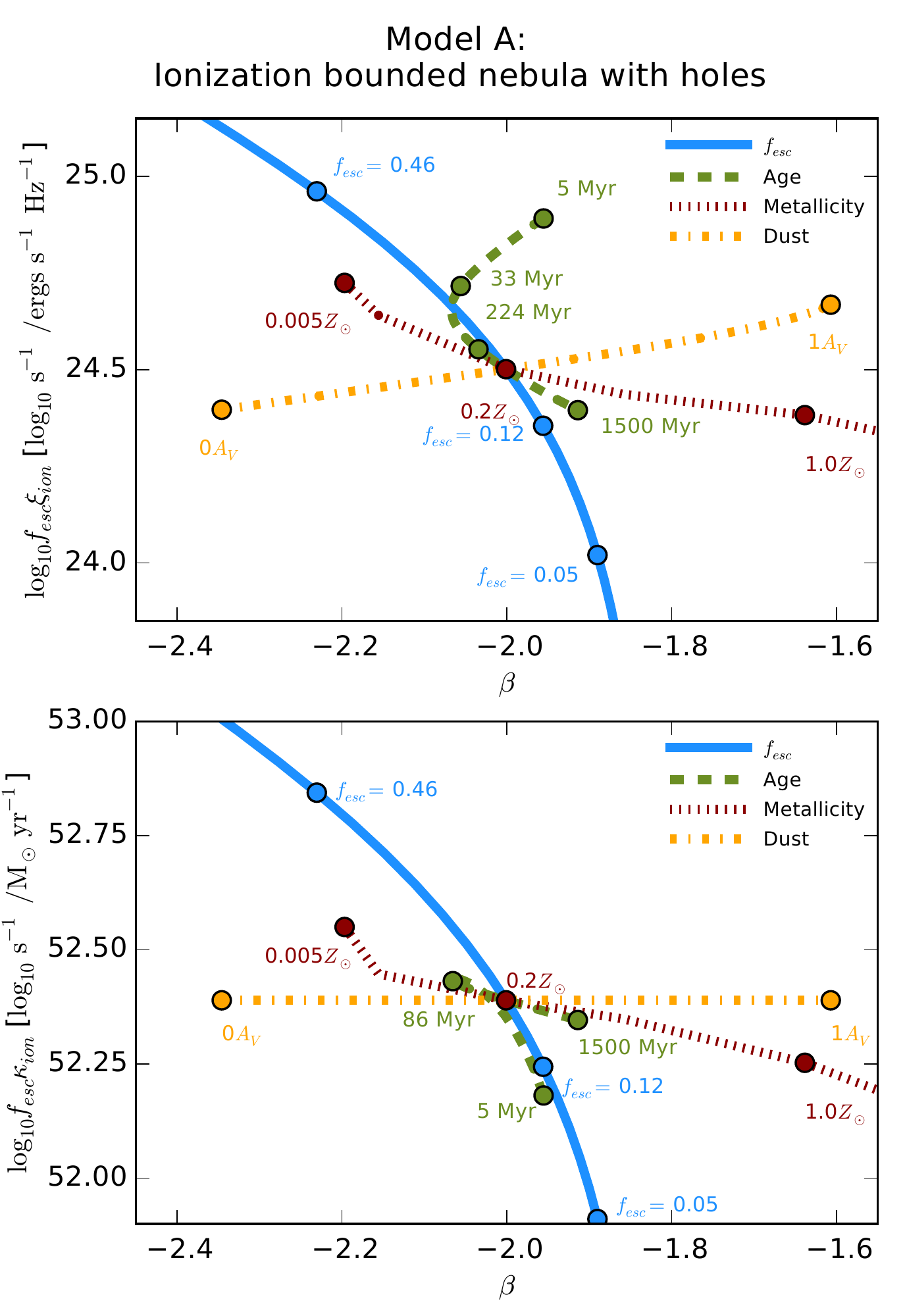}\quad
  \includegraphics[width=0.42\textwidth]{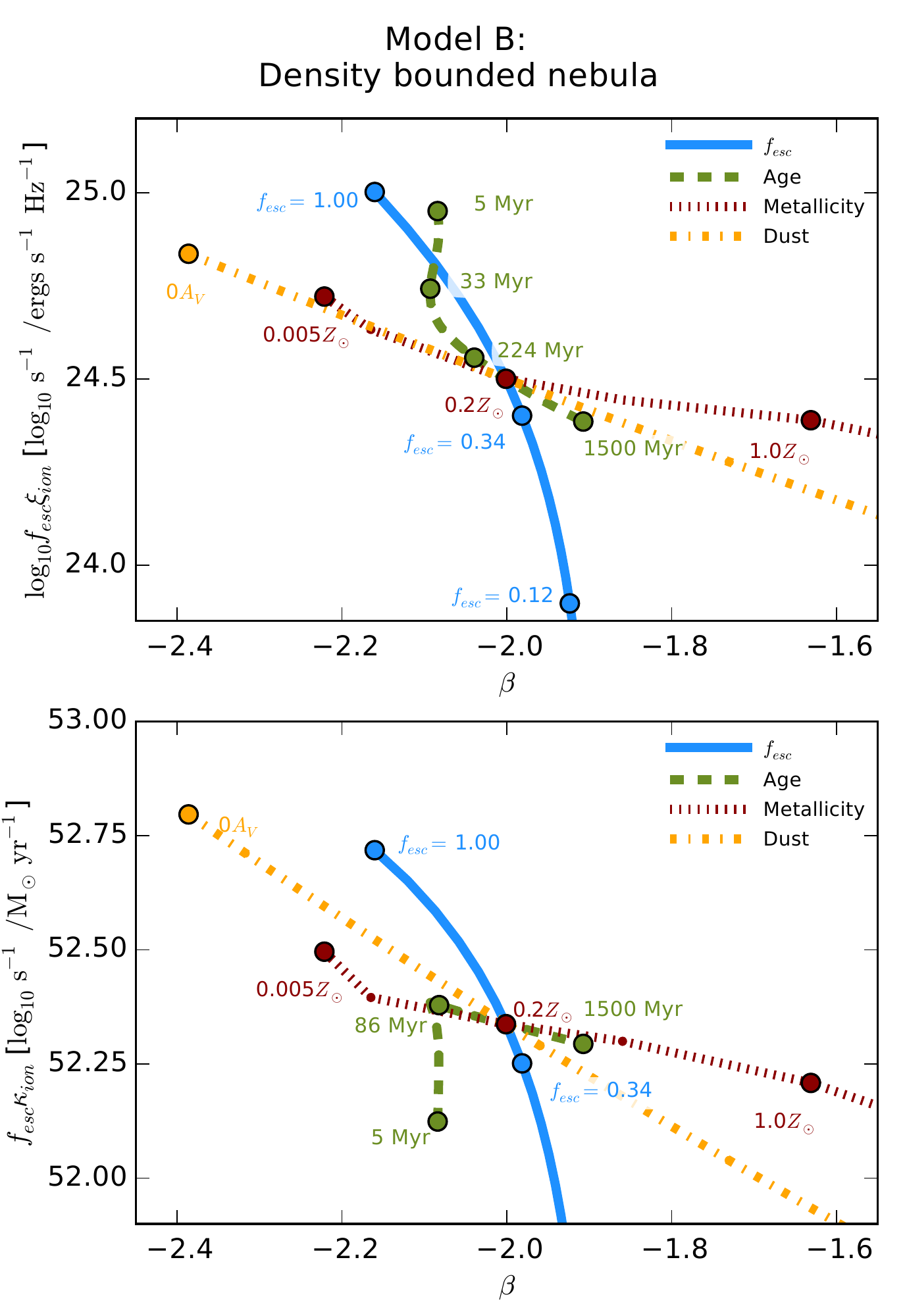}
  \caption{Evolution of $\log_{10}f_{\rm{esc}}\xi_{\rm{ion}}$ (top panels) and $\log_{10}f_{\rm{esc}}\kappa_{\rm{ion}}$ (bottom panels) vs $\beta$ as a function of changes in the stellar population age (green dashed), metallicity (red dotted), dust extinction (yellow dot-dashed) and escape fraction (blue continuous) with the remaining parameters fixed to the fiducial values listed in Table~\ref{tab:fiducial}. Values are plotted for both the ionization bounded nebula with holes (Model A; left panels) and density bounded nebula (Model B; right panels). Some individual points are labelled for both Lyman continuum escape models to illustrate the range and differences in evolution between each model. Note that the small difference in $\beta$ between Model A and Model B for the case of zero-dust ($0 A_{V}$) is due to the difference in nebular emission contribution for the two models; $f_{esc,neb} = 0.16$ and $f_{esc,neb} = 0.42$ for models A and B respectively.}
  \label{fig:xi_evol}
\end{figure*}

In the left panel of Fig.~\ref{fig:xi_evol}, we show how $\beta$ and $f_{\rm{esc}}\xi_{\rm{ion}}$ or $f_{\rm{esc}}\kappa_{\rm{ion}}$ vary as a function of each model parameter for the ionization-bounded nebula model (Model A) with the remaining parameters kept fixed at our fiducial $\beta = -2$ model (Table~\ref{tab:fiducial}). The corresponding values for the density bounded nebula model (Model B) are shown in the right-hand panel of Fig.~\ref{fig:xi_evol}. How $f_{\rm{esc}}\xi_{\rm{ion}}$ and $f_{\rm{esc}}\kappa_{\rm{ion}}$ vary with respect to $\beta$ for changes in the different parameters differs significantly:

\begin{itemize}
    \item \emph{Dust:} As has already been seen in Fig.~\ref{fig:fesc_cover}, $\beta$ varies strongly as a function of dust attenuation strength for the ionization-bounded nebula model. There is however minimal variation in $f_{\rm{esc}}\xi_{\rm{ion}}$ or $f_{\rm{esc}}\kappa_{\rm{ion}}$ with respect to that large change in $\beta$. The inferred $f_{\rm{esc}}\xi_{\rm{ion}}$ or $f_{\rm{esc}}\kappa_{\rm{ion}}$ can justifiably be considered constant as a function of redshift for this escape model if it is assumed that evolution in the dust extinction is responsible for the observed evolution of $\beta$.
    
    For model B, the density-bounded nebula, the large evolution in $\beta$ is coupled to a significant evolution in the inferred emissivity coefficients. For a change in the UV slope of $\Delta\beta \approx 0.2$, there is a corresponding evolution in $f_{\rm{esc}}\xi_{\rm{ion}}$ and$f_{\rm{esc}}\kappa_{\rm{ion}}$ of $\approx 0.19$ and $0.25$ dex respectively.
        
    \item \emph{Metallicity:} Between extremely sub-solar ($Z=0.005~Z_{\odot}$) and super-solar ($Z>1~Z_{\odot}$) metallicities, the variation in $\beta$ is large enough to account for the wide range of observed average $\beta$'s for both Lyman escape mechanisms. In this regard, metallicity evolution is a plausible mechanism to explain the apparent variation in $\beta$. However, such a wide variation in metallicities is not supported by current observations (see Section~\ref{sec:metallicity}) .
    
    Both $f_{\rm{esc}}\xi_{\rm{ion}}$ and $f_{\rm{esc}}\kappa_{\rm{ion}}$ vary by a factor of $\sim2$ across the full metallicity range modelled in this work, with bluer low-metallicity stellar populations producing more Lyman continuum photons per unit SFR/UV luminosity.
    
    \item \emph{Age:} Due to the strong nebular continuum contribution to the overall spectra at young ages, evolution in the stellar population age results in a more complicated $\beta$-emissivity relation. When nebular emission is included the continuum emission reddens the slope at very young ages before turning over at $t\approx100$ Myr and reddening with age towards ages of $t = 1$ Gyr and greater. For both LyC escape models, younger stellar populations results in a higher number of ionizing photons per unit UV luminosity. 

    For Model A, the effect of reddening by nebular emission at young ages is more pronounced due to the lower nebular region escape fraction ($f_{esc,neb} = 0.21$) in our fiducial model. The result of this reddening is that for the same $\beta = -2$, the corresponding $\log_{10} f_{\rm{esc}}\xi_{\rm{ion}}$ can be either $\approx 24.5$ or $\approx 25.$. This represents a potentially huge degeneracy if the ages of galaxy stellar populations are not well constrained.
    
    Between $z\sim 7$ and $z\sim4$, the evolution in stellar population age can only account for a reddening of $\Delta\beta \approx 0.1$, less than the evolution in both the normalisation of the colour-magnitude relation and in $\left \langle \beta  \right \rangle_{\rho_{\rm{UV}}}$. Similarly, it worth noting that at $z\sim7$, the assumption of a later onset for star-formation (e.g. $z\sim 9$, \citet{Collaboration:2015tp}) results in a $\sim0.14$ dex increase in the intrinsic $\xi_{\rm{ion}}$ and a change in the UV slope of $\Delta\beta \approx -0.065$. 
    
    \item \emph{$f_{\rm{esc}}$:} For both models of LyC escape, variation in $f_{\rm{esc}}$ has a strong evolution in $f_{\rm{esc}}\xi_{\rm{ion}}$ or$f_{\rm{esc}}\kappa_{\rm{ion}}$ with respect to changes in $\beta$. However, for Model B (density bounded nebula) the range of $\beta$ covered by the range of $f_{esc,neb}$ ($0 \leq f_{esc,neb} \leq 1$) is only $\approx 0.2$ dex, significantly less than the range of colours reached by variation in the other stellar population parameters.
\end{itemize}

\section{Estimated galaxy emissivity during reionization}\label{sec:results}
Using our improved understanding of the ionizing efficiencies of galaxies during the EoR, we can now estimate the total ionizing emissivity $\dot{N}_{\rm{ion}}$ of the galaxy population at high redshift following the prescription outlined in Equations~\ref{eq:Nion_UV} and \ref{eq:Nion_SFR}. We quantify $\rho_{\rm{UV}}$ and $\rho_{\rm{SFR}}$ using the latest available observations of the galaxy population extending deep in the epoch of reionization.

\subsection{Observations}\label{sec:observations}
Thanks to the deep and wide near-infrared observations of the CANDELS survey \citep{2011ApJS..197...35G,Koekemoer:2011br} and the extremely deep but narrow UDF12 survey \citep{Koekemoer:2013db}, there now exist direct constraints on the observed luminosity function deep into the epoch of reionization. In this paper, we will make use of the UV luminosity functions calculated by \citet{McLure:2013hh} and \citet{Schenker:2013cl} at $z = 7 - 9$ as part of the UDF12 survey along with the recent results of \citet{Bouwens:2014tx} and \citep{Finkelstein:2014ub} at $z \geq 4$ and the latest results from lensing clusters at $z >8$ \citep{Oesch:2014cs,McLeod:2014wz}.

A second, complimentary constraint on the amount of star-formation at high redshift is the stellar mass function and the integrated stellar mass density observed in subsequent epochs. As the time integral of all past star-formation, the stellar mass density can in principal be used to constrain the past star-formation rate if the star-formation history is known \citep{Stark:2007gi}.
A potential advantage of using the star-formation rate density in this manner is that by being able to probe further down the mass function, it may be possible to indirectly measure more star-formation than is directly observable at higher redshifts. Or to outline in other terms, if the total stellar mass density of all galaxies can be well known at $z\sim4$ or $z\sim5$, strict upper limits can be placed on the amount of unobserved (e.g. below the limiting depth of $z\sim8$ observations) or obscured star-formation at $z>6$.

At its simplest, the relation between a star-formation history, $S(t)$, and the resulting stellar mass  $M_{*}$ (or stellar mass density $\rho_{*}$) is given by
\begin{equation}
    M_{*}(t_{z}) = (1 - \epsilon_{z}) \times  \int_{t_{f}}^{t_{z}} S(t)~dt
\end{equation}
where $t_f$ and $t_z$ are the age of the Universe at the onset of star-formation and observed redshift respectively and $\epsilon_{z}$ is the fraction of mass returned to the ISM at the observed redshift. For a parametrised star-formation history, $F(t)$, which is normalised such that \( \int_{t_{f}}^{t_{z}} F(t)~dt = 1 \rm{M}_{\odot} (\rm{Mpc}^{-3}) \), we can substitute $S(t) = C_{obs,z}\times F(t)$. The normalisation, $C_{obs,z}$, accounts for the normalisation of the star-formation history required to match the observed stellar mass (or stellar mass density) at the redshift $z$. 

From recent observations of the stellar-mass density at $z > 4$, such as those from the stellar mass functions presented in \citet{Duncan:2014gh} and \citet{Grazian:2014vx}, it is then straight-forward to calculate the corresponding $C_{obs,z}$ and the inferred star-formation history ($C_{obs,z}\times F(t)$) for any assumed parametrisation. Motivated by the discussion of star-formation histories in Section~\ref{sec:assumptions}, we assume a normalised star-formation history which is $F(t) \propto t^{1.4}$ \citep{Salmon:2014tm}, as is done for the modelled $\beta$ values.

Using our stellar population models, we calculate $\epsilon$ for the star-formation history and metallicity used in this work at any desired redshift. We find that for a \citet{Chabrier:2003ki} IMF, $\epsilon_{z}$ varies from $\epsilon_{z} \approx 0.29$ at $z = 8$ to $\epsilon_{z} \approx 0.35$ at $z = 4$ (for a Salpeter IMF and constant SFH at $\sim 10$ Gyr old, we calculate $\epsilon = 0.279$ in agreement with the figure stated in \citetalias{Robertson:2013ji}).

\subsection{$\dot{N}_{\rm{ion}}$ for constant $f_{\rm{esc}}\xi_{\rm{ion}}$ and $f_{\rm{esc}}\kappa_{\rm{ion}}$}
\begin{figure*}
  \includegraphics[width=0.9\textwidth]{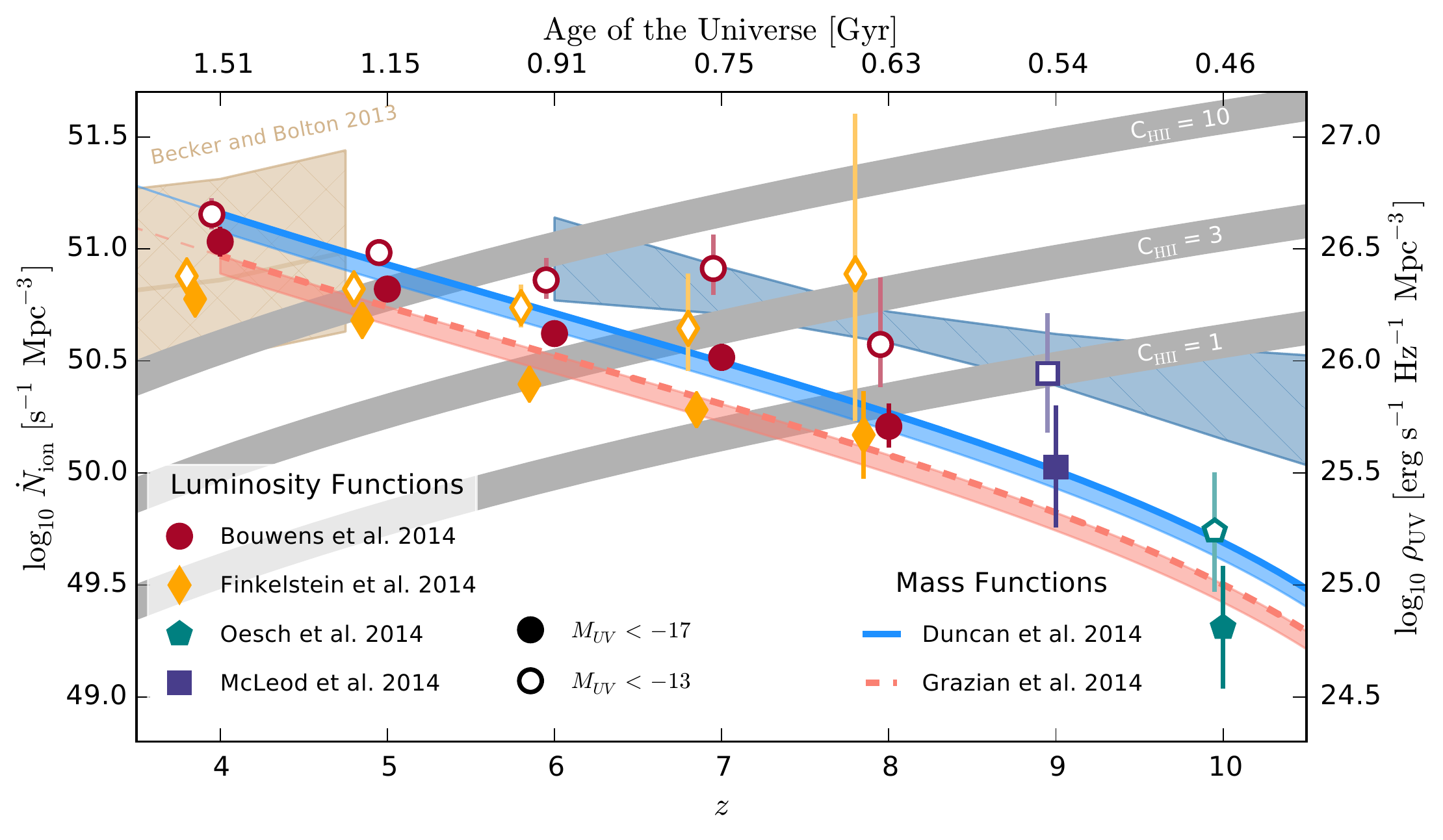}
  \caption{Ionizing emissivity $\dot{N}_{\rm{ion}}$ predicted for a fixed $f_{\rm{esc}}\xi_{\rm{ion}}$ for the UV luminosity function observations of \citet{Bouwens:2014tx}, \citet{Oesch:2014cs}, \citet{Finkelstein:2014ub} and \citet{McLeod:2014wz}. Filled symbols are for luminosity functions integrated down to $M_{\rm{UV}} = -17$ (typical limiting depth of  while open symbols correspond to a UV luminosity density integrated down to a constant $M_{\rm{UV}} = -13$ for all redshifts. Also shown are the ionizing emissivities inferred by the $z\sim4$ stellar mass density observations of \citet{Duncan:2014gh} (solid blue line) and \citet{Grazian:2014vx} (dashed pink line) for the power-law star-formation history  outlined in Section~\ref{sec:assumptions}. For the stellar mass function predictions, the thick solid and dashed line correspond to a constant $\log_{10}f_{\rm{esc}}\kappa_{\rm{ion}} = 52.39$ (Lyman escape model A) while the filled region shows the systematic offset when assuming $\log_{10}f_{\rm{esc}}\kappa_{\rm{ion}} = 52.34$ for Lyman escape model B, see the text for details. We show the IGM emissivity measurements and corresponding total errors of \citet{Becker:2013hc} (tan line and cross-hatched region respectively) and the ionizing background constraints of \citet{Bouwens:2015tm} (blue-gray diagonal-hatched region). The ionizing emissivity required to maintain reionization as a function of redshift and clumping factors ($C_{\textsc{Hii}}$) of one, three and ten are shown by the wide grey regions (\citeauthor{Madau:1999kl}~\citeyear{Madau:1999kl}, also see Equation~18 of \citeauthor{Bolton:2007gc}~\citeyear{Bolton:2007gc}).}
  \label{fig:Nion_constant}
\end{figure*}

In Fig.~\ref{fig:Nion_constant}, we show the estimated ionizing emissivity as a function of redshift for UV luminosity function observations of \citet{Bouwens:2014tx,Finkelstein:2014ub,Oesch:2014cs} and \citet{McLeod:2014wz}, assuming our fiducial constant $\log_{10}f_{\rm{esc}}\xi_{\rm{ion}} = 24.5$ (Table~\ref{tab:fiducial}, as assumed in \citeauthor{Robertson:2013ji}, \citeyear{Robertson:2013ji}). The integrated luminosity density and corresponding confidence intervals for each UV LF observation are estimated by drawing a set of LF parameters from the corresponding MCMC chain or likelihood distribution obtained in their fitting. This is repeated $10^{4}$ times to give a distribution from which we plot the median and $68\%$ confidence interval. The luminosity functions of \citet{Bouwens:2014tx} and \citet{Finkelstein:2014ub} span a large redshift range from $z\sim4$ to $z\sim8$ and predominantly make use of the same imaging data (including the ultra deep UDF12 observations, \citeauthor{Koekemoer:2013db}~\citeyear{Koekemoer:2013db}) but use different reductions of said data and differing selection and detection criteria, we refer the reader to the respective papers for more details. 

Also shown are the current constraints at $z\sim10$ based on the luminosity function from \citet{Oesch:2014cs}. Due to the small number of sources available at $z\sim10$ and the very large uncertainty in their redshift, the luminosity function is not well constrained at these redshifts. Fits using the \citet{Schechter:1976gl} parametrisation realistically allows only one free parameter to be varied while the remainder are fixed to their $z\sim8$ values (we plot the $\dot{N}_{\rm{ion}}$ predicted for the luminosity function parameters where $\phi_{*}$ is allowed to vary). Despite the large uncertainty, we include these values to illustrate the early suggestions of \citet{Oesch:2014cs} and other works \citep{Zheng:2012hu,Coe:2012ko} that the luminosity function (and hence the inferred underlying star-formation rate density) begins to fall more rapidly at $z \geq 9$ than extrapolations from lower redshift naively suggest. However, more recent analysis by \citet{McLeod:2014wz} of existing Frontier Fields data is in better agreement with the predicted luminosity density at $z \approx 9$. Completed observations of all six Frontier Fields clusters should provide significantly improved constraints at $z>9$ \citep{Coe:2014tq}, although the large cosmic variance of samples due to the volume effects of strong lensing will limit the constraints that can be placed at the highest redshifts \citep{Robertson:2014uv}.

For all redshift samples plotted in Fig.~\ref{fig:Nion_constant}, filled symbols correspond to the UV luminosity density integrated to $M_{\rm{UV}} = -17$ and $M_{\rm{UV}} = -13$ respectively. The predicted $\dot{N}_{\rm{ion}}$ for the UV luminosity functions of \citet{Schenker:2013cl} and \citet{McLure:2013hh} (not shown in Fig.~\ref{fig:Nion_constant}) effectively reproduce the UV luminosity density constraints outlined in \citetalias{Robertson:2013ji} and lie between those predicted by \citet{Bouwens:2014tx} and \citet{Finkelstein:2014ub}. The more recent works of \citet{Bouwens:2014tx} and \citet{Finkelstein:2014ub} show a greater disagreement between both themselves and previous works. This is a concern as it means that the choice of luminosity function (and hence the underlying selection/methodology) could have a significant effect on the conclusions drawn on galaxies' ability to complete or maintain reionization by the desired redshift.

At $z\sim6$, the UV luminosity density from galaxies brighter than the limiting depth observed by \citet{Bouwens:2014tx} is large enough to maintain reionization for a clumping factor of three. This is in contrast to the previous LF of \citet{Anonymous:96uKWdy6} and the results of \citet{Finkelstein:2014ub} which require a contribution from galaxies fainter than $M_{\rm{UV}} = -17$ (approximately the observational limits) to produce the $\dot{N}_{\rm{ion}}$ needed to maintain reionization. At $z\sim8$, the large uncertainties (and fitting degeneracies) in both the faint-end slope and characteristic luminosity of the luminosity function means that both of the luminosity density (and hence $\dot{N}_{\rm{ion}}$) assuming a constant $f_{\rm{esc}}\xi_{\rm{ion}}$) estimates included in this work agree within their $1\sigma$ errors. For the redshift dependent ionizing emissivity inferred by \citeauthor{Bouwens:2015tm} (\citeyear{Bouwens:2015tm} , assuming $C_{\textsc{Hii}} = 3$), faint galaxies down to at least $M_{\rm{UV}} = -13$ are required for all redshifts at $z \geq 6$ based on the current UV luminosity density constraints and the fiducial $f_{\rm{esc}}\xi_{\rm{ion}} = 24.5$.

To convert the star-formation rates inferred by the stellar mass densities observed by \citet{Duncan:2014gh} and \citet{Grazian:2014vx} to an $\dot{N}_{\rm{ion}}$ directly comparable with the LF estimates, we choose the $f_{\rm{esc}}\kappa_{\rm{ion}}$ at the $log_{10}f_{\rm{esc}}-A_{V}$ values where $\beta = -2$ and $\log_{10} f_{\rm{esc}}\xi_{\rm{ion}} = 24.5$. For the ionization bounded nebula with holes model, the corresponding $\log_{10} f_{\rm{esc}}\kappa_{\rm{ion}} =  52.39$, whilst for the density bounded nebula $\log_{10} f_{\rm{esc}}\kappa_{\rm{ion}} = 52.34$.

The ionizing photon rate predicted by the $z\sim4$ stellar mass functions of \citet{Duncan:2014gh} and \citet{Grazian:2014vx} for stellar masses greater than $10^{8.55}\/\rm{M}_{\odot}$ (the estimated lower limit from \citep{Duncan:2014gh} for which stellar masses can be reliably measured at $z\sim4$) are shown as the blue and pink lines plotted in Fig.~\ref{fig:Nion_constant}. Plotted as thick solid and dashed lines are the $\dot{N}_{\rm{ion}}$ assuming $\log_{10} f_{\rm{esc}}\kappa_{\rm{ion}} = 52.39$ with the corresponding shaded area illustrating the systematic offset for $\log_{10} f_{\rm{esc}}\kappa_{\rm{ion}} = 52.34$. 

The UV luminosity density at $z > 4$ implied by the SMD observations of \citet{Grazian:2014vx} (for $M>10^{8.55}\/\rm{M}_{\odot}$) are in good agreement with the UV LF estimates of \citet{Finkelstein:2014ub} when integrated down to $M_{\rm{UV}} = -17$ ($\approx$ observation limits). However, when integrating the stellar mass function down to significantly lower masses such as $> 10^{7} \rm{M}_{\odot}$, the shallower low-mass slope of \citep{Grazian:2014vx} results in a negligible increase in the total stellar mass density ($\approx 0.07$ dex). This is potentially inconsistent with the star-formation (and resulting stellar mass density) implied at $z > 6$ when galaxies from below the current observations limits of the luminosity functions are taken into account (open plotted symbols). 

The higher stellar mass density observed by \citet{Duncan:2014gh} results in $\dot{N}_{\rm{ion}}$ most consistent with those inferred by the \citet{Bouwens:2014tx} luminosity density ($M_{\rm{UV}} > -17$). Overall, there is reasonable agreement between the implied star-formation rates of the luminosity and stellar mass functions. However, both estimates have comparable systematic differences between different sets of observations.

Since the observational limit for the stellar mass function is currently limited by the reliability of stellar mass estimates for galaxies rather than their detection, better stellar mass estimates alone (through either deeper IRAC data or more informative fitting priors) could extend the observational limit for current high redshift galaxy samples. Improved constraints on the stellar mass functions at $z \leq 6$ are therefore a viable way of improving the constraints on SFR density and ionizing emissivity of faint galaxies at higher redshifts. 

\subsection{$\dot{N}_{\rm{ion}}$ for evolving $f_{\rm{esc}}\xi_{\rm{ion}}$}

To estimate what effect a $\beta$ dependent $f_{\rm{esc}}\xi_{\rm{ion}}$ or $f_{\rm{esc}}\kappa_{\rm{ion}}$ would have on the predicted $\dot{N}_{\rm{ion}}$, we assume two separate $f_{\rm{esc}}\xi_{\rm{ion}}(\beta)$ relations based on the predicted evolution of $f_{\rm{esc}}\xi_{\rm{ion}}$ and $f_{\rm{esc}}\kappa_{\rm{ion}}$ vs $\beta$ shown in Fig.~\ref{fig:xi_evol}. The first relation,  \emph{`ModelB\_dust'}, is based on the relatively shallow evolution of $f_{\rm{esc}}\xi_{\rm{ion}}$ and $f_{\rm{esc}}\kappa_{\rm{ion}}$ vs $\beta$ as a function of dust extinction for the density-bounded nebula model (Model B). Over the dynamic range in $\beta$, the \emph{ModelB\_dust} model evolves $\approx 0.5\/\rm{dex}$ from $\log_{10}f_{\rm{esc}}\xi_{\rm{ion}}(\beta = -2.3) \approx 24.75$ to $\log_{10}f_{\rm{esc}}\xi_{\rm{ion}}(\beta = -1.7) \approx 24.25$. 

The second relation, \emph{ModelA\_fesc}, follows the $f_{\rm{esc}}\xi_{\rm{ion}}$ or $f_{\rm{esc}}\kappa_{\rm{ion}}$ vs $\beta$ evolution as a function of $f_{\rm{esc}}$ for the ionization bounded nebula model (Model A). This model evolves from $\log_{10}f_{\rm{esc}}\xi_{\rm{ion}} \approx 25$ at $\beta = -2.3$ to effectively zero ionizing photons per unit luminosity/star-formation at $\beta = -1.8$. Due to the lack of constraints on $\beta$ for galaxies all the way down to $M_{\rm{UV}} = -13$, we set a lower limit on how steep the UV slope can become. This limit is chosen to match the UV slope for the dust-free, $f_{\rm{esc}} = 1$ scenario for fiducial model and has a slope of $\beta = -2.55$. We choose these two relations (three including the constant assumption above) because they correspond to the most likely mechanisms through which $\beta$ or the LyC escape fraction are expected to evolve.

Firstly, the constant (equivalent to \emph{`ModelA\_dust'}) and \emph{`ModelB\_dust'} models cover the assumption that evolution in the dust content of galaxies is responsible for the observed evolution in $\beta$ and any corresponding evolution in the ionizing efficiency of galaxies ($f_{\rm{esc}}\xi_{\rm{ion}}$ or $f_{\rm{esc}}\kappa_{\rm{ion}}$). Secondly, the \emph{`ModelA\_fesc'} model corresponds to an evolution in $f_{\rm{esc}}$ alone and for the ionization-bounded nebula with holes model, represents the steepest evolution of $f_{\rm{esc}}\xi_{\rm{ion}}$ (/$f_{\rm{esc}}\kappa_{\rm{ion}}$) with respect to $\beta$ of any of the parameters. While there is no obvious physical mechanism for such evolution at high-redshift, using this model we can at least link an inferred $f_{\rm{esc}}$ redshift evolution such as that from the \citet{Kuhlen:2012ka} and \citetalias{Robertson:2013ji} to a corresponding evolution in $\beta$ and vice-versa. In these works, the evolving escape fraction is parametrised as:
\begin{equation}\label{eq	:fesc_z_r13}
f_{\rm{esc}}(z) = f_{0} \times \left ( \frac{1 + z}{5} \right )^{\gamma }
\end{equation}
where $f_{0} = 0.054$ and $\gamma = 2.4$ \citepalias{Robertson:2013ji} and are constrained by the observed IGM emissivity values of \citet{FaucherGiguere:2008jc} at $z \leq 4$ and the WMAP total integrated optical depth measurements of \citet{Hinshaw:2013dd} at higher redshift. For the subsequent analysis we also show how the total ionizing emissivity would change following this evolution of $f_{\rm{esc}}$ (assuming a constant $\xi_{\rm{ion}} = 25.2$ in line with \citetalias{Robertson:2013ji}).

It is important to note here that the more recent measurements of the IGM emissivity at $z\sim4$ by \citet{Becker:2013hc} are a factor $\sim 2$ greater than those of \citet{FaucherGiguere:2008jc} at the same redshift. As such, the assumed values may under-estimate the zero-point $f_{0}$ and over-estimate the slope of the $f_{\rm{esc}}$ evolution compared to those fitted to the IGM emissivities of \citet{Becker:2013hc}. However, we include this to illustrate the effects that forcing consistency with IGM and optical depth measurements has on total ionizing emissivity for comparable underlying UV luminosity/star-formation rate density measurements relative to the assumption of a constant conversion.

 For the star-formation history assumed throughout this work, both of these models should in principle also take into account the reddening of the intrinsic UV slope due to age evolution (see Section~\ref{sec:beta_effect_stellar_pop}). However, we neglect this contribution  in the following analysis for two reasons. Firstly, the effect of age evolution on the observed $\beta$ is small in comparison to that of $f_{\rm{esc}}$ or dust for these models. Secondly, the vector relating $\beta$ and $f_{\rm{esc}}\xi_{\rm_{ion}}$ for increasing age is either almost parallel to or slightly steeper than those for $f_{\rm{esc}}$ and dust respectively. As such, the effects on the inferred $f_{\rm{esc}}\xi_{\rm_{ion}}$ will be negligible.

\subsubsection{Evolving luminosity-averaged $f_{\rm{esc}}\xi_{\rm{ion}}$}
To explore how a $\beta$ dependent coefficient would change the inferred emissivities, we first calculate a constant $f_{\rm{esc}}\xi_{\rm{ion}}$ for each redshift based on the luminosity weighted $\left \langle \beta  \right \rangle_{\rm{UV}}$. The $z\sim4$ to $z\sim7$ $\left \langle \beta  \right \rangle_{\rm{UV}}$ used are those from \citet{Bouwens:2013vf}, with the $z\sim8$ value based on the fit outlined in Equation~\ref{eq:beta_muv_fit}. 

\begin{figure}
  \includegraphics[width=0.48\textwidth]{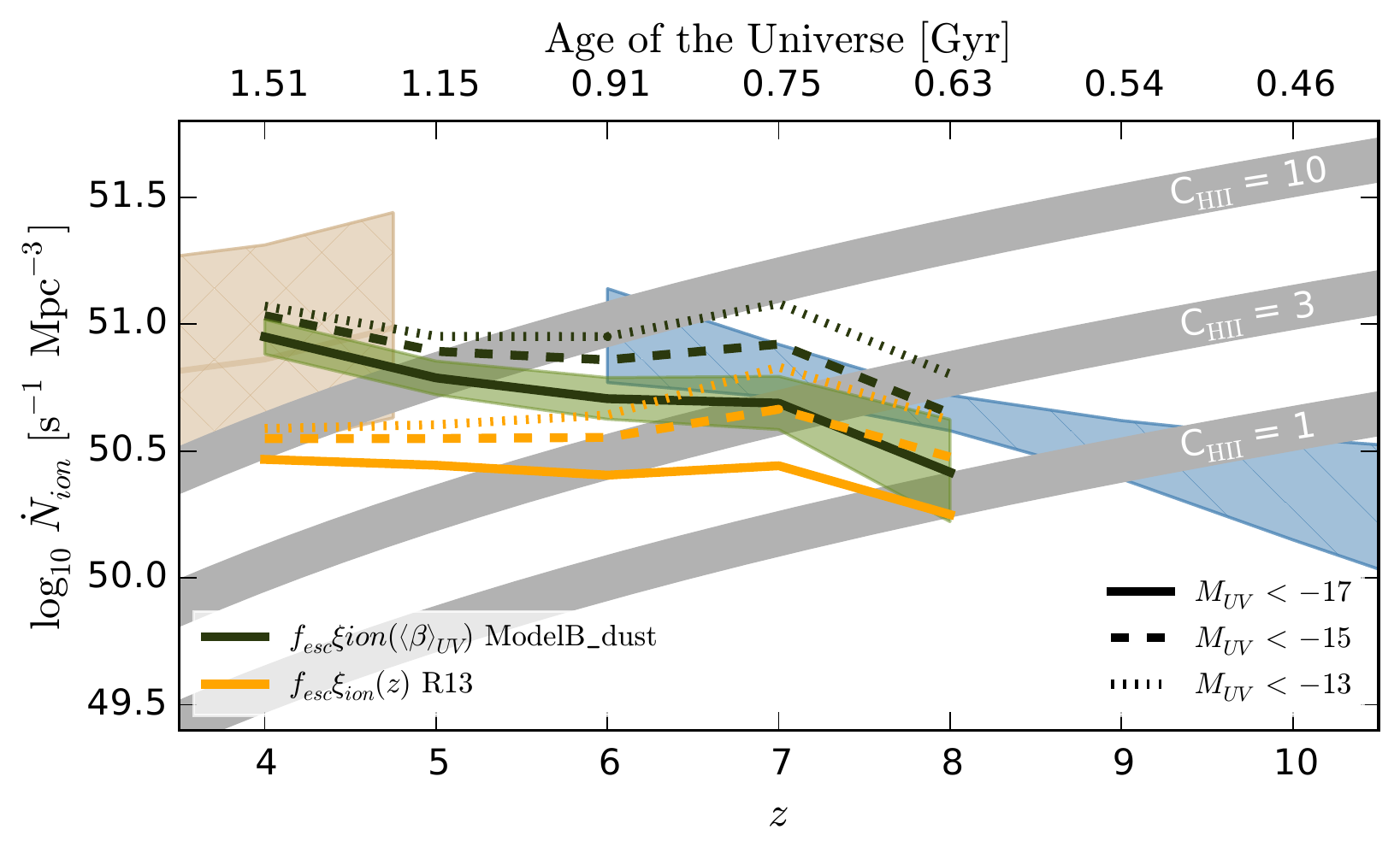}
  \includegraphics[width=0.48\textwidth]{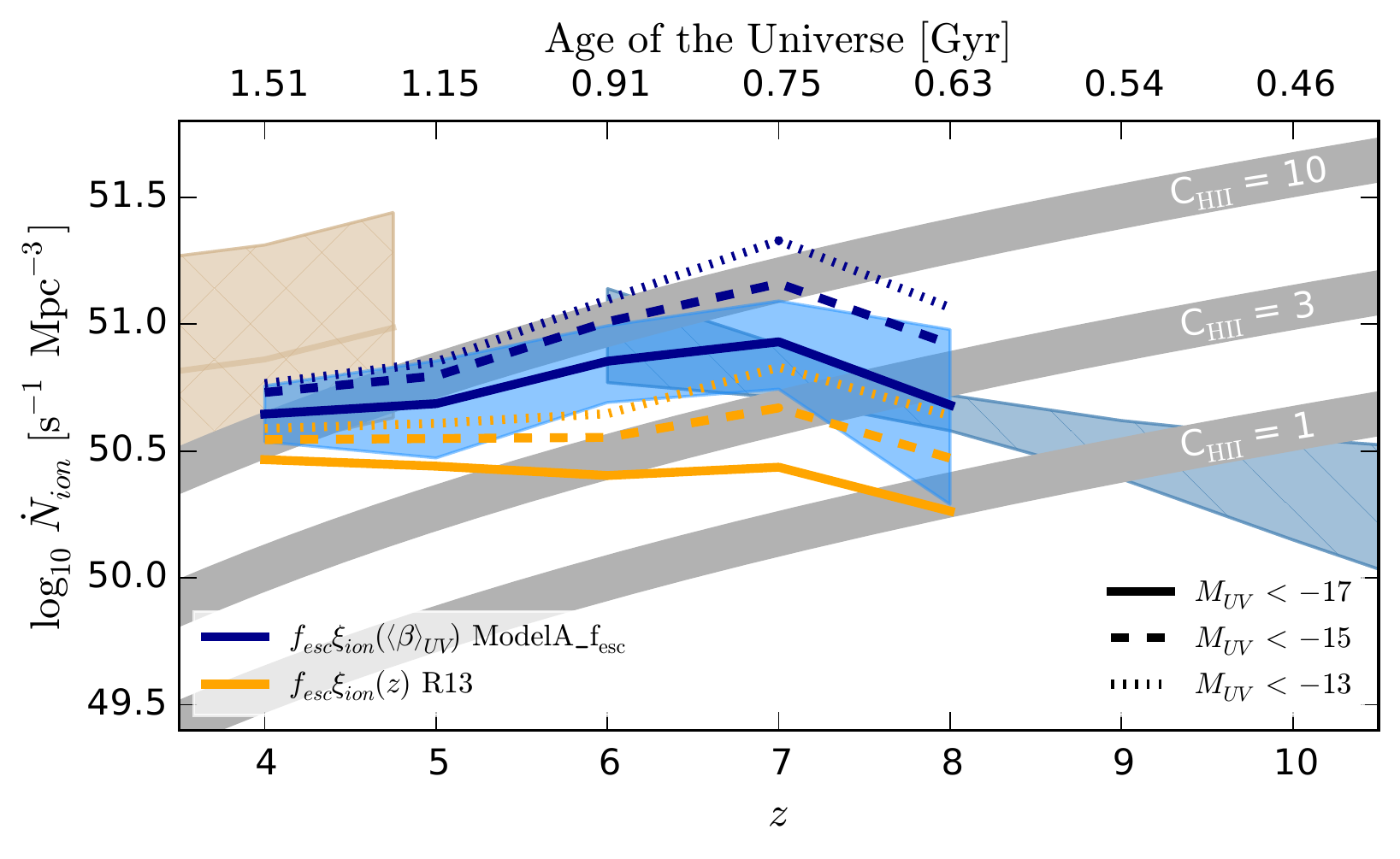}
  \caption{Ionizing emissivity, $\dot{N}_{\rm{ion}}$, predicted by the luminosity functions measured by \citet{Bouwens:2014tx} for an evolving $f_{\rm{esc}}\xi_{\rm{ion}}$ as a function of redshift, based on the luminosity weighted average $\beta$. See text for details on the assumed $f_{\rm{esc}}\xi_{\rm{ion}}$ as a function of $\left \langle \beta  \right \rangle_{\rm{UV}}$ for the \emph{ModelB\_dust} (top - green) and \emph{ModelA\_fesc} (bottom - blue) models. In both panels, the thick solid and dashed lines correspond to the UV luminosity density integrated down to the observational limit and a constant $M_{\rm{UV}} = -13$ respectively. The yellow lines indicate the emissivity predicted for redshift dependent $f_{\rm{esc}}$ as inferred by \citetalias{Robertson:2013ji}, see Equation~\ref{eq	:fesc_z_r13}. As in Figure~\ref{fig:Nion_constant}, we show the IGM emissivity measurements and corresponding total errors of \citet{Becker:2013hc} (tan line and cross-hatched region respectively) and the ionizing background constraints of \citet{Bouwens:2015tm} (blue-gray diagonal-hatched region).}
  \label{fig:Nion_xi_kappa_z}
\end{figure}

For clarity, we plot the resulting predicted emissivities only for the UV luminosity densities predicted by \citet{Bouwens:2014tx} luminosity function parametrisations, these are shown in Fig.~\ref{fig:Nion_xi_kappa_z} for both the \emph{ModelB\_dust} and \emph{ModelA\_fesc} (top and bottom panels respectively) $f_{\rm{esc}}\xi_{\rm{ion}}(\beta)$ relations. The shaded regions around the solid green and blue lines (top and bottom panels respectively) represent the uncertainty on $f_{\rm{esc}}\xi_{\rm{ion}}$ due to the statistical uncertainty in $\left \langle \beta  \right \rangle_{\rm{UV}}$. The full statistical uncertainties include the luminosity density errors illustrated in Fig.~\ref{fig:Nion_constant} and are included in Appendix Table~\ref{tab:fiducial}.

By assuming a $\beta$ dependent coefficient, the estimated $\dot{N}_{\rm{ion}}$ evolution changes shape to a much shallower evolution with redshift. For our \emph{ModelB\_dust} $\beta$ relation, the decline in $\dot{N}_{\rm{ion}}$ from $z=4$ to $z=8$ is reduced by $\approx 0.25$ dex when the luminosity function is integrated down to the limit of $M_{\rm{UV}} < -13$. The effect is even stronger for the \emph{ModelA\_fesc} $\beta$ evolution, with the $\dot{N}_{\rm{ion}}$ actually increasing over this redshift. The larger $f_{\rm{esc}}\xi_{\rm{ion}}$ inferred by the increasingly blue UV slopes at high redshift are able to balance the rapid decrease in UV luminosity density.

A key effect of the increasing ionizing efficiency with increasing redshifts is that for both $f_{\rm{esc}}\xi_{\rm{ion}}(\beta)$ relations explored here, the observable galaxy population at $z\sim7$ is now capable of maintaining reionization for a clumping factor of $C_{\textsc{Hii}} = 3$. This is in contrast to the result inferred when assuming a constant $f_{\rm{esc}}\xi_{\rm{ion}}$.  

However, we know the brightest galaxies are in fact also the reddest and that any of our predicted $f_{\rm{esc}}\xi_{\rm{ion}}(\beta)$ relations imply they are therefore the least efficient at producing ionizing photons. The application of an average $f_{\rm{esc}}\xi_{\rm{ion}}$ (even one weighted by the relative contributions to the luminosity density) may give a misleading impression of the relative contribution the brightest galaxies make to the ionizing background during the EoR. A more accurate picture can be obtained by applying a luminosity dependent $\beta$ relation (e.g. $f_{\rm{esc}}\xi_{\rm{ion}}(M_{\rm{UV}})$) to observed luminosity function and integrating the ionizing emissivity, $\dot{N}_{\rm{ion}}$, from this.

\subsubsection{Luminosity dependent $f_{\rm{esc}}\xi_{\rm{ion}}$}
\begin{figure}
  \includegraphics[width=0.48\textwidth]{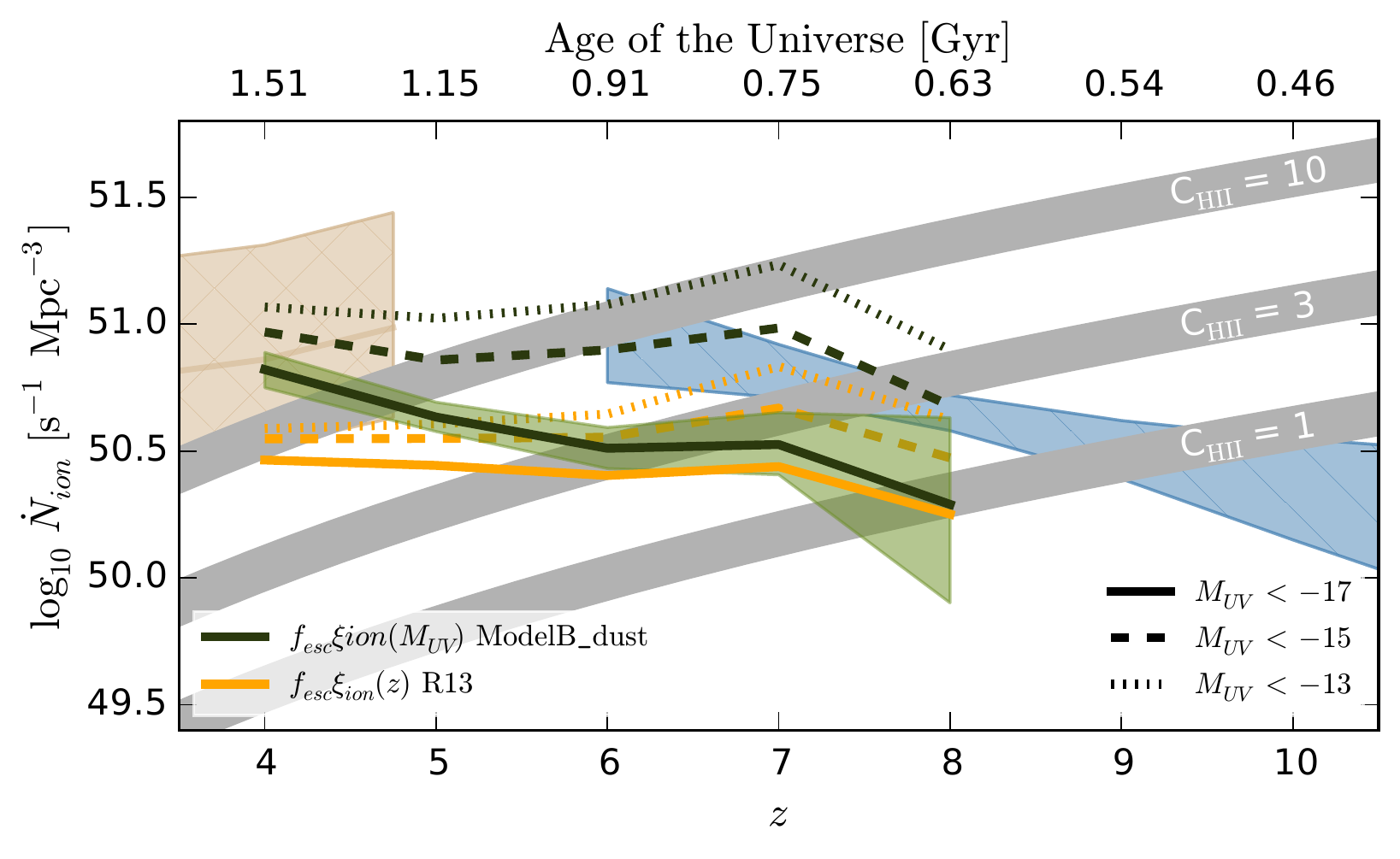}
  \includegraphics[width=0.48\textwidth]{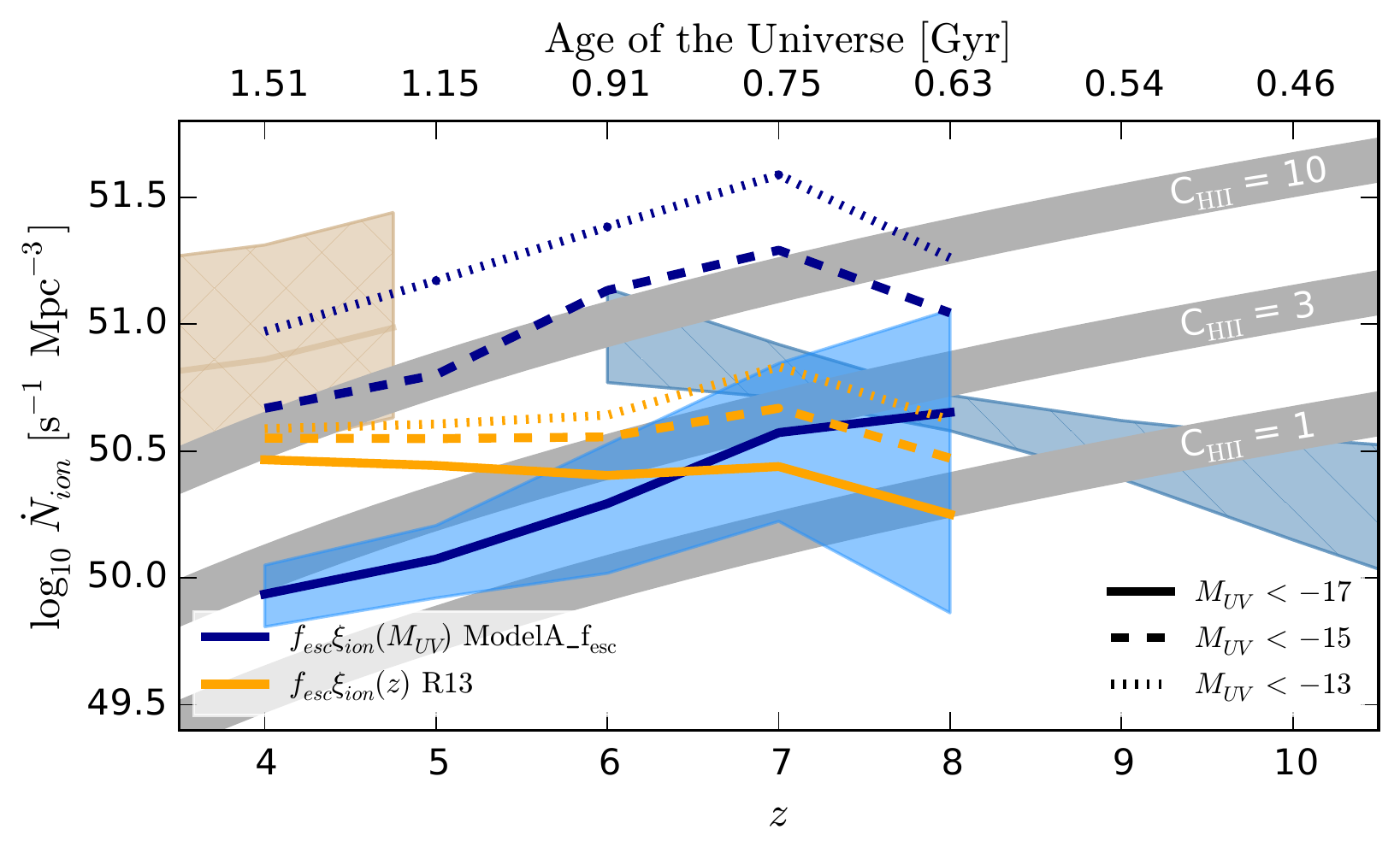}
  \caption{Ionizing emissivity, $\dot{N}_{\rm{ion}}$, predicted by the luminosity functions measured by \citet{Bouwens:2014tx} for a luminosity dependent $f_{\rm{esc}}\xi_{\rm{ion}}$. See text for details on the assumed $f_{\rm{esc}}\xi_{\rm{ion}}$ as a function of $\beta(M_{\rm{UV}})$ for the \emph{ModelB\_dust}(top) and \emph{ModelA\_fesc} (bottom) models. In both panels, the thick solid, dashed and dotted lines correspond to the UV luminosity density integrated down to $M_{\rm{UV}} = -17$, -15 and -13 respectively. The yellow lines indicate the emissivity predicted for redshift dependent $f_{\rm{esc}}$ as inferred by \citetalias{Robertson:2013ji}, see Equation~\ref{eq	:fesc_z_r13}. As in Figure~\ref{fig:Nion_constant}, we show the IGM emissivity measurements and corresponding total errors of \citet{Becker:2013hc} (tan line and cross-hatched region respectively) and the ionizing background constraints of \citet{Bouwens:2015tm} (blue-gray diagonal-hatched region).}
  \label{fig:Nion_xi_kappa_zMuv}
\end{figure}

Using the observed $\beta(M_{\rm{UV}})$ relations of \citet{Bouwens:2013vf} (Fig.~\ref{fig:beta_weightedavg}) and our models for $f_{\rm{esc}}\xi_{\rm{ion}}(\beta)$, we next calculate $\dot{N}_{\rm{ion}}$ a function of both the changing luminosity function and the evolving colour magnitude relation. Fig.~\ref{fig:Nion_xi_kappa_zMuv} shows the evolution of $\dot{N}_{\rm{ion}}$ based on these assumptions, again for the \citet{Bouwens:2014tx} luminosity function parametrisations. For both the \emph{ModelB\_dust} and \emph{ModelA\_fesc} relations, the emissivity of galaxies above the limiting depths of the observations are reduced due to the fact that the brighter galaxies have significantly redder observed $\beta$s. 

In the case of the \emph{ModelA\_fesc} $\beta$ evolution and the \emph{ModelB\_dust} relation at high redshift, the difference between the $\dot{N}_{\rm{ion}}$ for $M_{\rm{UV}} < -17$ and $M_{\rm{UV}} < -13$ is quite significant. This is due to the observed steepening of the colour-magnitude relation (Section~\ref{sec:betas}) at higher redshifts results in a larger $f_{\rm{esc}}\xi_{\rm{ion}}$ evolution between the brightest and faintest galaxies in the luminosity function. When integrating the UV luminosity function from fainter magnitudes, the number of ionizing photons produced per unit UV luminosity density significantly increases. At lower redshifts, the sharp decline in $f_{\rm{esc}}\xi_{\rm{ion}}$ for redder galaxies in \emph{ModelA\_fesc} means that the brightest galaxies can potentially contribute very little to the total ionizing emissivity.

The total galaxy ionizing emissivity ($M_{\rm{UV}} < -13$) for both models of $\beta$ evolution is high enough to maintain reionization (for a clumping factor of $C_{\textsc{Hii}} \lesssim 3$) at all redshifts. In fact, only galaxies down to a rest-frame magnitude of $M_{\rm{UV}} = -15$ are required to match these rates. Furthermore, despite the increased $\dot{N}_{\rm{ion}}$ predicted at $z > 5$ both models are in good agreement with the observed IGM emissivities of \citet{Becker:2013hc} at lower redshifts when the luminosity functions are integrated down to a limit of $M_{\rm{UV}} < -13$. Including galaxies such faint galaxies does however result in a potential over-production of ionizing photons at $z> 6$ based on the emissivity estimates of \citet{Bouwens:2015tm}. Given that we have shown that significantly lower values of $f_{\rm{esc}}$ are still consistent with the observed colours (as stated in Section~\ref{sec:slp}), an overproduction for the optimistic models assumed does not necessarily present an immediate problem.

From these results we can see that changes in the ionizing efficiency of galaxies during EoR which are still consistent with the evolving UV continuum slopes have significantly less effect on the predicted total ionizing emissivity ($M_{\rm{UV}} < -13$) at $z\sim4$ than at higher redshifts based on the current observations. This is a crucial outcome with regards to current numerical models for the epoch of reionization, as it allows for a wider range of reionization histories which are still consistent with both the observed UV luminosity/SFR-density and lower redshift IGM emissivity estimates. 

\section{Discussion and future prospects}\label{sec:discussion}
In several previous studies of the reionization history of the Universe the conclusion has been drawn that at earlier times in the epoch of reionization, galaxies must have been more efficient at ionizing the surrounding IGM than similar galaxies at lower redshift \citep{Becker:2013hc,Kuhlen:2012ka,Robertson:2013ji}. Based on the constraints on galaxy stellar populations and escape fractions explored in Section~\ref{sec:models} and their application to the existing observations in Section~\ref{sec:results}, it is not yet possible to establish that one particular galaxy property is evolving to cause such an increase in ionizing efficiency. 

However, what we find in this work is that evolution in galaxy properties, such as dust extinction and escape fraction (or some combination of these and others), which are consistent with the observed colour evolution of high-redshift galaxies can readily account for any increase in the ionizing efficiency required by other constraints such as the total optical depth. Ongoing and future observations of both local and distant galaxies will be able to provide much tighter constraints on the evolving galaxy properties.

If the observed $\beta$ evolution is a result of dust alone, as is assumed by \citet{2012ApJ...754...83B} and other works, the inferred evolution in $f_{\rm{esc}}\xi_{\rm{ion}}$ as a function of $\beta$ is highly dependent on the assumed model of Lyman continuum escape and hence the underlying geometry of dust and gas. For example, if the channels through which LyC photons escape are dust-free (as in Model A),  the effect of the dust evolution will have negligible effect on the emissivity of galaxies as a function of $\beta$ (as discussed in Section~\ref{sec:beta_effect_stellar_pop}). Real galaxies will of course be significantly more complicated (and messy) than the simple toy models adopted in this work, as such the channels through which LyC escape may also contain significant quantities of dust.  We find that if we modify Model A such that the dust screen is extended to include the low-density channels, the resulting model is indistinguishable from Model B (density bounded nebula) with regards to $\beta$ as a function of $f_{\rm{esc}}$ or $A_{V}$. Such a model closely matches that observed by \citet{Borthakur:2014bz} (see also \citeauthor{Heckman:2011ju}~\citeyear{Heckman:2011ju}) for a local analog of galaxies during the EoR and represents our best model for LyC escape. 

For any of the plausible causes for the luminosity and redshift evolution of $\beta$ (dust, metallicity, escape fraction), the models explored in this work infer that fainter/low-mass galaxies are emitting more ionizing photons per unit star-formation into the IGM than their higher mass counterparts. Currently, simulations of galaxies at high redshift draw somewhat differing conclusions on the mass/luminosity dependence of the escape fraction. Based on a combination of theoretical models and the existing limited observations, \citet{Gnedin:2008ib} find that angular averaged escape fraction increases with higher star-formation rates and galaxy masses, the inverse of what we predict based on $\beta$ alone. However, in isolation from the model predictions, the observational data explored by \citet{Gnedin:2008ib} does not place any strong constaints on the luminosity dependence of $f_{\rm{esc}}$ \citep{Giallongo:2002hi,FernandezSoto:2003ev,Shapley:2006cq}.

Subsequent simulations predict the opposite luminosity dependence, in better agreement with the colour evolution predictions of this work \citep{Razoumov:2010bh,Yajima:2010fb}. Recent work exploring the escape fraction of both typical \citep{Kimm:2014gv} and dwarf \citep{Wise:2014kt} galaxies at $z\geq7$ find that the instantaneous escape fraction (measured at the virial radius in these simulations) is inversely proportional to the halo mass. However, as discussed by \citet{Kimm:2014gv}, the average instantaneous escape fraction may be somewhat misleading due to the bursty nature of star-formation in their models and the delay between the peak SFR and maximum escape fraction for an episode of star-formation. They find that the time-averaged escape fraction weighted by the overall LyC photon production rate remains roughly constant for size haloes. Improved measurements on the stellar or halo mass dependence of $f_{\rm{esc}}$ are therefore clearly crucial.

Although direct measurements of the LyC escape fraction for galaxies during EoR will never be possible due to the effects of IGM absorption along the line of sight, measurements of the escape fraction as a function of stellar mass and luminosity (or SFR) at $z \lesssim 3$ should soon be possible due to the deep UV imaging of new surveys such as the UVUDF \citep{Teplitz:2013jg} and the forthcoming GOODS UV Legacy Survey (PI: Oesch, GO13872). The wealth of ancillary data available in these fields (both photometric and spectroscopic) should make it possible to tightly constrain $f_{\rm{esc}}$, $\xi_{\rm{ion}}$ or $\kappa_{\rm{ion}}$, and $\beta$ for either individual galaxies or samples stacked by galaxy properties. Measuring $\beta$ vs $f_{\rm{esc}}\xi_{\rm{ion}}$ at $2 \lesssim z \lesssim 3$ would significantly reduce systematic errors in the inferred $\dot{N}_{\rm{ion}}$ during the EoR from incorrect or poorly informed assumptions on $f_{\rm{esc}}\xi_{\rm{ion}}$.

Given the large degeneracies in $\beta$ with respect to the various stellar population parameters, understanding  $\beta$ vs $f_{\rm{esc}}\xi_{\rm{ion}}$ both at $z\sim3$ and during the EoR will require an improved understanding of what is responsible for the observed $\beta$ evolution. With ALMA observations of statistically significant samples of galaxies at high-redshift, it should be possible to make strong constraints on not just the strength and attenuation curve of the dust extinction, but also the location and geometry of the dust relative to the gas and star-formation within galaxies \citep{DeBreuck:2014eo}.

With the new generation of near-infrared sensitive spectrographs allowing precision spectroscopic measurement of metallicities and dust out to $z >3$ (e.g. MOSFIRE: \citet{Kriek:2014uw}), it will be possible to place much more accurate priors on the expected ages, metallicities and star-formation histories for galaxies during the EoR. Finally, as with many outstanding problems in astrophysics, the launch of the James Webb Space Telescope will address many of the systematic and statistical uncertainties which limit current observations. Crucially, JWST should be able to probe much fainter galaxy populations, potentially down to rest-frame magnitudes of $M_{\rm{UV}} = -15$ and below. Based on the findings in this paper, such observations might even mean we are finally able to observe the full galaxy population responsible for powering reionization.

\section{Summary}\label{sec:summary}
In this work, we explore in-depth the ionizing photon budget of galaxies during the epoch of reionization based solely on the observed galaxy properties. For the latest observational constraints on the star-formation rate and UV luminosity density at $z > 4$, we assess the ionizing emissivity consistent with new constraints on the rest-frame UV colours of galaxies at these redshifts.

Using a comprehensive set of SED models for two plausible Lyman continuum escape mechanisms -- previously outlined in \citet{Zackrisson:2013iz} -- we explore in detail the relationship between the UV continuum slope $\beta$ and the number of ionizing photons produced per unit UV luminosity or star-formation ($\xi_{\rm{ion}}$ and $\kappa_{\rm{ion}}$ respectively). We find that the ionizing efficiencies assumed by several previous works ($\log_{10} f_{\rm{esc}}\xi_{\rm{ion}} = 24.5-24.6$: \citet{Robertson:2013ji,Kuhlen:2012ka}) are still consistent with the current $\beta$ observations during the EoR. However, for both of the LyC escape models explored here, this assumption is close to the maximum efficiency which is still consistent with the fiducial UV slope typically considered at these redshifts ($\beta = -2$). Based on our SED modelling, escape fractions or ionizing efficiency which are $1~\rm{dex}$ lower than typically assumed are still consistent with the observed galaxy colours.

Applying the fiducial $\log_{10} f_{\rm{esc}}\xi_{\rm{ion}} = 24.5$ to the the latest observations of the luminosity and mass functions at $z\geq4$, we find that at $z\sim6$, the observed population can produce enough ionizing photons to maintain reionization assuming a clumping factor $C_{\textsc{Hii}} = 3$. At earlier times, we confirm previous results which found that galaxies from below our current observation limits are required to produce enough ionizing photons to maintain reionization at $z\sim7$ and beyond \citep{2010Natur.468...49R,Robertson:2013ji,Kuhlen:2012ka,Finkelstein:2012hr}.

Motivated by increasing evidence for a luminosity dependence of the UV continuum slope  \citep{Bouwens:2013vf,Rogers:2014bn} and evidence for evolution in this relation with redshift \citep{Bouwens:2013vf}, we explore the effects of assuming an ionizing efficiency which is not constant but varies with the observed $\beta$. The two galaxy properties that are able to plausibly account for the required range of observed $\beta$'s, dust extinction and $f_{\rm{esc}}$, both predict an ionizing efficiency which increases for increasingly blue UV continuum slopes. While the other galaxy properties such as age and metallicity predict similar trends, current observations do not support a large enough variation to account for the required range of observed UV slopes.

We find that when assuming an ionizing efficiency based on the luminosity-weighted average $\beta$, the currently observable galaxy population alone is now able to maintain reionization at $z\sim7$ (assuming $C_{\textsc{Hii}} = 3$). Despite this increase in efficiency at early times, the predicted $\dot{N}_{\rm{ion}}$ at $z < 5$ remain consistent with measurements based on the IGM \citep{Becker:2013hc}.

Assuming instead that the ionizing efficiency of galaxies is dependent on their luminosity, the observed $M_{\rm{UV}} - \beta$ relations and our SED models can result in significant changes in the inferred ionizing photon budget. Since our models suggest that redder (brighter) galaxies have lower $f_{\rm{esc}}\xi_{\rm{ion}}$ than their blue (faint) counterparts, the inferred ionizing photon budget for the currently observable galaxy population may be significantly reduced, especially at lower redshifts. However, because of the increasing importance of faint galaxies (which have higher inferred $f_{\rm{esc}}\xi_{\rm{ion}}$), only galaxies down to $M_{\rm{UV}} \approx -15$ may be required to produce the required ionizing photons.

Our conclusion is that the inferred ability of galaxies to complete or maintain reionization is highly dependent on the stellar population assumptions used to predict their ionizing efficiencies. Crucially though, the models explored in this study can potentially allow for a wide range of reionization histories whilst remaining consistent with the observed colour evolution and luminosity (or star-formation rate) density during this epoch. Future work on constraining both the colour and luminosity dependence of $f_{\rm{esc}}$ at lower redshifts as well as measuring the ages and dust content of galaxies during the EoR will be vital in understanding the precise ionizing emissivity of galaxies throughout this epoch.

\section*{Acknowledgements}
We thank the referee for their feedback and help in improving the paper.  KD also thanks James Bolton for conversations and feedback throughout the writing process as well as Rychard Bouwens and Steven Finkelstein for sharing the fitting data for their respective papers. Finally, we would like to acknowledge funding from the Science and Technology Facilities Council (STFC) and the Leverhulme Trust.

\footnotesize{
    \bibliography{bibtex_lib}
}

\appendix

\section{Data Tables}\label{app:tables}
In this appendix we present the estimated $\rho_{\rm{UV}}$ and $\dot{N}_{\rm{ion}}$ at $4 \leq z \leq 8$ for the range of assumptions outlined in Section~\ref{sec:results}.  we list the calculated properties for integration limits $M_{\rm{UV}} =$ -17, -15 and -13 respectively. In Table~\ref{tab:bouwens_N} we list the values based on the luminosity function parametrisations of \citet{Bouwens:2014tx} and are plotted in Figures~\ref{fig:Nion_constant}, \ref{fig:Nion_xi_kappa_z} and \ref{fig:Nion_xi_kappa_zMuv}. In Table~\ref{tab:finkelstein_N}, we list the corresponding values for the luminosity function parametrisations of \citep{Finkelstein:2014ub}.
For both sets of values, we include errors based on the uncertainties in the luminosity function parameters and the random errors in the weighted average of $\beta$ or the best-fit $\beta-M_{\rm{UV}}$ slope parameters \citep{Bouwens:2013vf} as appropriate. When assuming an SMC-like dust attenuation \citep{Pei:1992ey}, to match the fiducial $\beta = -2$ and $\log_{10}f_{\rm{esc}} \xi_{\rm{ion}}=24.5$ for LyC escape Model B, we find $f_{\rm{esc}} = 0.28$ and $A_{V} = 0.08$ are required. 

\begin{table*}\label{tab:bouwens_N}
  \caption{Calculated values of $\rho_{\rm{UV}}$ and $\dot{N}_{\rm{ion}}$ for the different integration limits and efficiency assumptions explored in the paper, based on the luminosity function parametrisations of \citet{Bouwens:2014tx}. For each calculated value, we include statistical errors from the uncertainties in the \citet{Schechter:1976gl} parameters and $\beta$ observations. Also shown are the effects of some of the assumptions made in Section~\ref{sec:assumptions} and their corresponding systematic changes to the estimated values.}
  \begin{tabular}{cc|cccccc}
  \hline
   & Limit ($M_{\rm{UV}}$) & $z\sim4$ & $z\sim5$ & $z\sim6$ & $z\sim7$ & $z\sim8$ & \\
   \hline
  $\log_{10} \rho_{\rm{UV}}$ 							& -17 & \input{EoR_data/EoRtable_Bouwens_rho.txt} \\
   (erg s$^{-1}$ Hz$^{-1}$ Mpc$^{-3}$)	& -15 & \input{EoR_data/EoRtable_Bouwens_rho_m15.txt} \\
																& -13 & \input{EoR_data/EoRtable_Bouwens_rho_m13.txt} \\

	 \multicolumn{7}{c}{}\\
	 \multicolumn{7}{c}{$\log_{10}f_{\rm{esc}}\xi_{\rm{ion}} = 24.5$}\\\hline
$\log_{10} \dot{N}_{\rm{ion}}$	 		& -17 & \input{EoR_data/EoRtable_Bouwens.txt} \\
(s$^{-1}$ Mpc$^{-3}$) 				& -15 & \input{EoR_data/EoRtable_Bouwens_m15.txt} \\
												& -13 & \input{EoR_data/EoRtable_Bouwens_m13.txt} \\
												
	 \multicolumn{7}{c}{}\\
	 \multicolumn{7}{c}{$\log_{10}f_{\rm{esc}}\xi_{\rm{ion}} \propto \left \langle \beta  \right \rangle_{\rho_{\rm{UV}}}(z)$}\\ \hline
	
	  & & \multicolumn{5}{c}{$ModelB\_Dust$}\\
	  
$\log_{10} \dot{N}_{\rm{ion}}$	 		& -17 & \input{EoR_data/EoRtable_Bouwens_z_shal.txt} \\
(s$^{-1}$ Mpc$^{-3}$) 				& -15 & \input{EoR_data/EoRtable_Bouwens_z_shal_m15.txt} \\
												& -13 & \input{EoR_data/EoRtable_Bouwens_z_shal_m13.txt} \\
												
	  & & \multicolumn{5}{c}{$ModelA\_f_{\rm{esc}}$}\\
	  
$\log_{10} \dot{N}_{\rm{ion}}$	 		& -17 & \input{EoR_data/EoRtable_Bouwens_z.txt} \\
(s$^{-1}$ Mpc$^{-3}$) 				& -15 & \input{EoR_data/EoRtable_Bouwens_z_m15.txt} \\
												& -13 & \input{EoR_data/EoRtable_Bouwens_z_m13.txt} \\											

	 \multicolumn{7}{c}{}\\
	 \multicolumn{7}{c}{$\log_{10}f_{\rm{esc}}\xi_{\rm{ion}}(M_{\rm{UV}})$}\\\hline
	 
	  & & \multicolumn{5}{c}{$ModelB\_Dust$}\\
	  
$\log_{10} \dot{N}_{\rm{ion}}$	 		& -17 & \input{EoR_data/EoRtable_Bouwens_zMuv_shal.txt} \\
(s$^{-1}$ Mpc$^{-3}$) 				& -15 & \input{EoR_data/EoRtable_Bouwens_zMuv_shal_m15.txt} \\
												& -13 & \input{EoR_data/EoRtable_Bouwens_zMuv_shal_m13.txt} \\
												
	  & & \multicolumn{5}{c}{$ModelA\_f_{\rm{esc}}$}\\
	  
$\log_{10} \dot{N}_{\rm{ion}}$	 		& -17 & \input{EoR_data/EoRtable_Bouwens_zMuv.txt} \\
(s$^{-1}$ Mpc$^{-3}$) 				& -15 & \input{EoR_data/EoRtable_Bouwens_zMuv_m15.txt} \\
												& -13 & \input{EoR_data/EoRtable_Bouwens_zMuv_m13.txt} \\		

\end{tabular}
 \begin{tabular}{cc|cccccc}
	 \multicolumn{7}{c}{}\\
	 \multicolumn{7}{c}{Systematic Uncertainties}\\\hline
	 \multicolumn{2}{c}{Salpeter IMF} & \multicolumn{5}{c}{$\Delta log_{10}\kappa_{\rm{ion}} = -0.19$}\\

	 \multicolumn{2}{c}{Dust: Calzetti w/ SMC-like extrapolation} &  \multicolumn{2}{c}{Model A: $\Delta log_{10}\xi_{\rm{ion}} = 0$} & \multicolumn{3}{c}{Model B: $\Delta log_{10}\xi_{\rm{ion}} = +0.10$}\\
	 \multicolumn{2}{c}{} &  \multicolumn{2}{c}{} & \multicolumn{3}{c}{ (For fiducial values)}\\
 
	 	 \multicolumn{2}{c}{Dust: SMC \citep{Pei:1992ey}} &  \multicolumn{2}{c}{Model A: $\Delta log_{10}\xi_{\rm{ion}} = 0$} & \multicolumn{3}{c}{Model B: $\Delta log_{10}\xi_{\rm{ion}} = +0.18$} \\
	 	 \multicolumn{2}{c}{} &  \multicolumn{2}{c}{} & \multicolumn{3}{c}{(For fiducial $f_{\rm{esc}} = 0.42$, $A_{V} = 0.08$, $\beta = -2$)}

  \end{tabular}
\end{table*}

\begin{table*}\label{tab:finkelstein_N}
  \caption{Calculated values of $\rho_{\rm{UV}}$ and $\dot{N}_{\rm{ion}}$ for the different integration limits and efficiency assumptions explored in the paper, based on the luminosity function parametrisations of \citet{Finkelstein:2014ub}. For each calculated value, we include statistical errors from the uncertainties in the \citet{Schechter:1976gl} parameters and $\beta$ observations. Also shown are the effects of some of the assumptions made in Section~\ref{sec:assumptions} and their corresponding systematic changes to the estimated values.}
  \begin{tabular}{cc|cccccc}
   \hline
   & Limit ($M_{\rm{UV}}$) & $z\sim4$ & $z\sim5$ & $z\sim6$ & $z\sim7$ & $z\sim8$ \\
   \hline
  $\log_{10} \rho_{\rm{UV}}$ 							& -17 & \input{EoR_data/EoRtable_Fink_rho.txt} \\
   (erg s$^{-1}$ Hz$^{-1}$ Mpc$^{-3}$)	& -15 & \input{EoR_data/EoRtable_Fink_rho_m15.txt} \\
																& -13 & \input{EoR_data/EoRtable_Fink_rho_m13.txt} \\

	 \multicolumn{7}{c}{}\\
	 \multicolumn{7}{c}{$\log_{10}f_{\rm{esc}}\xi_{\rm{ion}} = 24.5$}\\\hline
$\log_{10} \dot{N}_{\rm{ion}}$	 		& -17 & \input{EoR_data/EoRtable_Fink.txt} \\
(s$^{-1}$ Mpc$^{-3}$) 				& -15 & \input{EoR_data/EoRtable_Fink_m15.txt} \\
												& -13 & \input{EoR_data/EoRtable_Fink_m13.txt} \\
												
	 \multicolumn{7}{c}{}\\
	 \multicolumn{7}{c}{$\log_{10}f_{\rm{esc}}\xi_{\rm{ion}} \propto \left \langle \beta  \right \rangle_{\rho_{\rm{UV}}}(z)$}\\ \hline
	
	  & & \multicolumn{5}{c}{$ModelB\_Dust$}\\
	  
$\log_{10} \dot{N}_{\rm{ion}}$	 		& -17 & \input{EoR_data/EoRtable_Fink_z_shal.txt} \\
(s$^{-1}$ Mpc$^{-3}$) 				& -15 & \input{EoR_data/EoRtable_Fink_z_shal_m15.txt} \\
												& -13 & \input{EoR_data/EoRtable_Fink_z_shal_m13.txt} \\
												
	  & & \multicolumn{5}{c}{$ModelA\_f_{\rm{esc}}$}\\
	  
$\log_{10} \dot{N}_{\rm{ion}}$	 		& -17 & \input{EoR_data/EoRtable_Fink_z.txt} \\
(s$^{-1}$ Mpc$^{-3}$) 				& -15 & \input{EoR_data/EoRtable_Fink_z_m15.txt} \\
												& -13 & \input{EoR_data/EoRtable_Fink_z_m13.txt} \\											

	 \multicolumn{7}{c}{}\\
	 \multicolumn{7}{c}{$\log_{10}f_{\rm{esc}}\xi_{\rm{ion}}(M_{\rm{UV}})$}\\\hline
	 
	  & & \multicolumn{5}{c}{$ModelB\_Dust$}\\
	  
$\log_{10} \dot{N}_{\rm{ion}}$	 		& -17 & \input{EoR_data/EoRtable_Fink_zMuv_shal.txt} \\
(s$^{-1}$ Mpc$^{-3}$) 				& -15 & \input{EoR_data/EoRtable_Fink_zMuv_shal_m15.txt} \\
												& -13 & \input{EoR_data/EoRtable_Fink_zMuv_shal_m13.txt} \\
												
	  & & \multicolumn{5}{c}{$ModelA\_f_{\rm{esc}}$}\\
	  
$\log_{10} \dot{N}_{\rm{ion}}$	 		& -17 & \input{EoR_data/EoRtable_Fink_zMuv.txt} \\
(s$^{-1}$ Mpc$^{-3}$) 				& -15 & \input{EoR_data/EoRtable_Fink_zMuv_m15.txt} \\
												& -13 & \input{EoR_data/EoRtable_Fink_zMuv_m13.txt} \\		

\end{tabular}
 
 \begin{tabular}{cc|cccccc}
	 \multicolumn{7}{c}{}\\
	 \multicolumn{7}{c}{Systematic Uncertainties}\\\hline
	 \multicolumn{2}{c}{Salpeter IMF} & \multicolumn{5}{c}{$\Delta log_{10}\kappa_{\rm{ion}} = -0.19$}\\

	 \multicolumn{2}{c}{Dust: Calzetti w/ SMC-like extrapolation} &  \multicolumn{2}{c}{Model A: $\Delta log_{10}\xi_{\rm{ion}} = 0$} & \multicolumn{3}{c}{Model B: $\Delta log_{10}\xi_{\rm{ion}} = +0.10$}\\
	 \multicolumn{2}{c}{} &  \multicolumn{2}{c}{} & \multicolumn{3}{c}{ (For fiducial values)}\\
 
	 	 \multicolumn{2}{c}{Dust: SMC \citep{Pei:1992ey}} &  \multicolumn{2}{c}{Model A: $\Delta log_{10}\xi_{\rm{ion}} = 0$} & \multicolumn{3}{c}{Model B: $\Delta log_{10}\xi_{\rm{ion}} = +0.18$} \\
	 	 \multicolumn{2}{c}{} &  \multicolumn{2}{c}{} & \multicolumn{3}{c}{(For fiducial $f_{\rm{esc}} = 0.42$, $A_{V} = 0.08$, $\beta = -2$)} \\
\end{tabular}

\end{table*}

\label{lastpage}

\end{document}

%% file: EoR_data/EoRtable_Bouwens_rho.txt
$26.53^{+0.07}_{-0.07}$ & $26.32^{+0.05}_{-0.05}$ & $26.12^{+0.06}_{-0.06}$ & $26.02^{+0.06}_{-0.06}$ & $25.71^{+0.10}_{-0.10}$ & 

%% file: EoR_data/EoRtable_Bouwens_rho_m15.txt
$26.62^{+0.07}_{-0.07}$ & $26.43^{+0.05}_{-0.06}$ & $26.27^{+0.07}_{-0.07}$ & $26.25^{+0.10}_{-0.08}$ & $25.92^{+0.19}_{-0.13}$ & 

%% file: EoR_data/EoRtable_Bouwens_rho_m13.txt
$26.65^{+0.07}_{-0.07}$ & $26.48^{+0.06}_{-0.06}$ & $26.36^{+0.10}_{-0.08}$ & $26.41^{+0.15}_{-0.12}$ & $26.07^{+0.30}_{-0.19}$ & 

%% file: EoR_data/EoRtable_Bouwens.txt
$51.03^{+0.07}_{-0.07}$ & $50.82^{+0.05}_{-0.05}$ & $50.62^{+0.06}_{-0.06}$ & $50.52^{+0.06}_{-0.06}$ & $50.21^{+0.10}_{-0.10}$ & 

%% file: EoR_data/EoRtable_Bouwens_m15.txt
$51.12^{+0.07}_{-0.07}$  & $50.93^{+0.05}_{-0.06}$  & $50.77^{+0.07}_{-0.07}$  & $50.75^{+0.10}_{-0.08}$  & $50.42^{+0.19}_{-0.13}$  & 

%% file: EoR_data/EoRtable_Bouwens_m13.txt
$51.15^{+0.07}_{-0.07}$  & $50.98^{+0.06}_{-0.06}$  & $50.86^{+0.10}_{-0.08}$  & $50.91^{+0.15}_{-0.12}$  & $50.57^{+0.30}_{-0.19}$  & 

%% file: EoR_data/EoRtable_Bouwens_z_shal.txt
$50.95^{+0.07}_{-0.07}$ & $50.79^{+0.07}_{-0.07}$ & $50.71^{+0.08}_{-0.08}$ & $50.69^{+0.10}_{-0.10}$ & $50.42^{+0.20}_{-0.19}$ & 

%% file: EoR_data/EoRtable_Bouwens_z_shal_m15.txt
$51.03^{+0.07}_{-0.07}$ & $50.89^{+0.07}_{-0.07}$ & $50.86^{+0.09}_{-0.09}$ & $50.92^{+0.13}_{-0.12}$ & $50.64^{+0.26}_{-0.22}$ & 

%% file: EoR_data/EoRtable_Bouwens_z_shal_m13.txt
$51.07^{+0.07}_{-0.07}$ & $50.95^{+0.07}_{-0.07}$ & $50.95^{+0.11}_{-0.10}$ & $51.09^{+0.18}_{-0.15}$ & $50.80^{+0.32}_{-0.28}$ & 

%% file: EoR_data/EoRtable_Bouwens_z.txt
$50.65^{+0.11}_{-0.11}$ & $50.69^{+0.17}_{-0.21}$ & $50.85^{+0.14}_{-0.16}$ & $50.93^{+0.16}_{-0.19}$ & $50.68^{+0.30}_{-0.39}$ & 

%% file: EoR_data/EoRtable_Bouwens_z_m15.txt
$50.73^{+0.11}_{-0.11}$ & $50.80^{+0.16}_{-0.21}$ & $51.01^{+0.15}_{-0.17}$ & $51.16^{+0.18}_{-0.20}$ & $50.92^{+0.32}_{-0.39}$ & 

%% file: EoR_data/EoRtable_Bouwens_z_m13.txt
$50.77^{+0.11}_{-0.11}$ & $50.85^{+0.17}_{-0.22}$ & $51.10^{+0.16}_{-0.17}$ & $51.33^{+0.21}_{-0.22}$ & $51.06^{+0.39}_{-0.43}$ & 

%% file: EoR_data/EoRtable_Bouwens_zMuv_shal.txt
$50.82^{+0.07}_{-0.07}$ & $50.63^{+0.06}_{-0.06}$ & $50.51^{+0.08}_{-0.08}$ & $50.52^{+0.12}_{-0.12}$ & $50.29^{+0.34}_{-0.38}$ & 

%% file: EoR_data/EoRtable_Bouwens_zMuv_shal_m15.txt
$50.97^{+0.07}_{-0.07}$ & $50.86^{+0.07}_{-0.06}$ & $50.90^{+0.11}_{-0.11}$ & $50.98^{+0.15}_{-0.15}$ & $50.67^{+0.28}_{-0.41}$ & 

%% file: EoR_data/EoRtable_Bouwens_zMuv_shal_m13.txt
$51.07^{+0.07}_{-0.07}$ & $51.02^{+0.08}_{-0.08}$ & $51.08^{+0.14}_{-0.13}$ & $51.24^{+0.19}_{-0.17}$ & $50.89^{+0.35}_{-0.37}$ & 

%% file: EoR_data/EoRtable_Bouwens_zMuv.txt
$49.94^{+0.11}_{-0.13}$ & $50.07^{+0.13}_{-0.15}$ & $50.29^{+0.23}_{-0.27}$ & $50.57^{+0.27}_{-0.35}$ & $50.65^{+0.40}_{-0.79}$ & 

%% file: EoR_data/EoRtable_Bouwens_zMuv_m15.txt
$50.67^{+0.09}_{-0.09}$ & $50.80^{+0.13}_{-0.11}$ & $51.13^{+0.15}_{-0.16}$ & $51.29^{+0.18}_{-0.22}$ & $51.04^{+0.30}_{-0.76}$ & 

%% file: EoR_data/EoRtable_Bouwens_zMuv_m13.txt
$50.97^{+0.11}_{-0.11}$ & $51.17^{+0.12}_{-0.13}$ & $51.38^{+0.17}_{-0.16}$ & $51.59^{+0.22}_{-0.21}$ & $51.26^{+0.37}_{-0.53}$ & 

%% file: EoR_data/EoRtable_Fink_rho.txt
$26.28^{+0.01}_{-0.01}$ & $26.18^{+0.01}_{-0.01}$ & $25.90^{+0.02}_{-0.02}$ & $25.78^{+0.06}_{-0.06}$ & $25.67^{+0.19}_{-0.19}$ & 

%% file: EoR_data/EoRtable_Fink_rho_m15.txt
$26.35^{+0.02}_{-0.02}$ & $26.28^{+0.02}_{-0.02}$ & $26.10^{+0.06}_{-0.05}$ & $26.00^{+0.15}_{-0.13}$ & $26.04^{+0.46}_{-0.40}$ & 

%% file: EoR_data/EoRtable_Fink_rho_m13.txt
$26.38^{+0.02}_{-0.02}$ & $26.32^{+0.03}_{-0.02}$ & $26.24^{+0.10}_{-0.09}$ & $26.15^{+0.25}_{-0.19}$ & $26.39^{+0.71}_{-0.66}$ & 

%% file: EoR_data/EoRtable_Fink.txt
$50.78^{+0.01}_{-0.01}$ & $50.68^{+0.01}_{-0.01}$ & $50.40^{+0.02}_{-0.02}$ & $50.28^{+0.06}_{-0.06}$ & $50.17^{+0.19}_{-0.19}$ & 

%% file: EoR_data/EoRtable_Fink_m15.txt
$50.85^{+0.02}_{-0.02}$ & $50.78^{+0.02}_{-0.02}$ & $50.60^{+0.06}_{-0.05}$ & $50.50^{+0.15}_{-0.13}$ & $50.54^{+0.46}_{-0.40}$ & 

%% file: EoR_data/EoRtable_Fink_m13.txt
$50.88^{+0.02}_{-0.02}$ & $50.82^{+0.03}_{-0.02}$ & $50.74^{+0.10}_{-0.09}$ & $50.65^{+0.25}_{-0.19}$ & $50.89^{+0.71}_{-0.66}$ & 

%% file: EoR_data/EoRtable_Fink_z_shal.txt
$50.69^{+0.02}_{-0.02}$ & $50.65^{+0.04}_{-0.04}$ & $50.48^{+0.06}_{-0.06}$ & $50.45^{+0.11}_{-0.10}$ & $50.38^{+0.25}_{-0.26}$ & 

%% file: EoR_data/EoRtable_Fink_z_shal_m15.txt
$50.77^{+0.02}_{-0.02}$ & $50.74^{+0.05}_{-0.05}$ & $50.68^{+0.09}_{-0.08}$ & $50.67^{+0.17}_{-0.15}$ & $50.78^{+0.46}_{-0.45}$ & 

%% file: EoR_data/EoRtable_Fink_z_shal_m13.txt
$50.79^{+0.03}_{-0.03}$ & $50.79^{+0.05}_{-0.05}$ & $50.83^{+0.12}_{-0.11}$ & $50.82^{+0.26}_{-0.21}$ & $51.12^{+0.72}_{-0.67}$ & 

%% file: EoR_data/EoRtable_Fink_z.txt
$50.39^{+0.09}_{-0.09}$ & $50.55^{+0.15}_{-0.21}$ & $50.63^{+0.13}_{-0.16}$ & $50.70^{+0.16}_{-0.19}$ & $50.63^{+0.35}_{-0.42}$ & 

%% file: EoR_data/EoRtable_Fink_z_m15.txt
$50.46^{+0.09}_{-0.09}$ & $50.65^{+0.15}_{-0.21}$ & $50.83^{+0.14}_{-0.17}$ & $50.91^{+0.21}_{-0.22}$ & $51.01^{+0.52}_{-0.54}$ & 

%% file: EoR_data/EoRtable_Fink_z_m13.txt
$50.49^{+0.09}_{-0.09}$ & $50.69^{+0.16}_{-0.21}$ & $50.97^{+0.17}_{-0.18}$ & $51.06^{+0.28}_{-0.27}$ & $51.34^{+0.78}_{-0.72}$ & 

%% file: EoR_data/EoRtable_Fink_zMuv_shal.txt
$50.56^{+0.02}_{-0.02}$ & $50.50^{+0.03}_{-0.03}$ & $50.31^{+0.07}_{-0.06}$ & $50.27^{+0.14}_{-0.13}$ & $50.27^{+0.38}_{-0.44}$ & 

%% file: EoR_data/EoRtable_Fink_zMuv_shal_m15.txt
$50.69^{+0.03}_{-0.03}$ & $50.70^{+0.04}_{-0.04}$ & $50.78^{+0.11}_{-0.11}$ & $50.71^{+0.22}_{-0.21}$ & $50.82^{+0.52}_{-0.55}$ & 

%% file: EoR_data/EoRtable_Fink_zMuv_shal_m13.txt
$50.77^{+0.04}_{-0.04}$ & $50.84^{+0.06}_{-0.05}$ & $51.03^{+0.15}_{-0.14}$ & $50.95^{+0.31}_{-0.29}$ & $51.26^{+0.76}_{-0.80}$ & 

%% file: EoR_data/EoRtable_Fink_zMuv.txt
$49.63^{+0.11}_{-0.13}$ & $49.92^{+0.13}_{-0.16}$ & $50.16^{+0.22}_{-0.25}$ & $50.30^{+0.30}_{-0.38}$ & $50.66^{+0.44}_{-0.75}$ & 

%% file: EoR_data/EoRtable_Fink_zMuv_m15.txt
$50.34^{+0.07}_{-0.08}$ & $50.61^{+0.11}_{-0.10}$ & $51.06^{+0.14}_{-0.16}$ & $51.00^{+0.26}_{-0.30}$ & $51.16^{+0.58}_{-0.75}$ & 

%% file: EoR_data/EoRtable_Fink_zMuv_m13.txt
$50.61^{+0.10}_{-0.10}$ & $50.95^{+0.11}_{-0.11}$ & $51.37^{+0.17}_{-0.17}$ & $51.29^{+0.35}_{-0.34}$ & $51.62^{+0.80}_{-0.87}$ & 